\documentclass[preprint,preprintnumbers,amsmath,amssymb,floatfix,nofootinbib]{revtex4}
\usepackage{graphicx}
\usepackage{slashed}
\usepackage{url}
\newcommand{\sss}{\scriptscriptstyle}
\begin{document}          
\preprint{BNL-HET-03/17}
\preprint{FSU-HEP-2003-1015}
\preprint{UB-HET-03/05}
\preprint{hep-ph/0311067}
\title{Exclusive Higgs Boson Production with bottom quarks at
  Hadron Colliders}
\author{S.~Dawson}
\email{dawson@quark.phy.bnl.gov}
\affiliation{Physics Department, Brookhaven National Laboratory,
Upton, NY 11973-5000, USA}
\author{C.~B.~Jackson}
\email{jackson@hep.fsu.edu}
\affiliation{Physics Department, Florida State University,
Tallahassee, FL 32306-4350, USA}
\author{L.~Reina}
\email{reina@hep.fsu.edu}
\affiliation{Physics Department, Florida State University,
Tallahassee, FL 32306-4350, USA}
\author{D.~Wackeroth}
\email{dow@ubpheno.physics.buffalo.edu}
\affiliation{Department of Physics, SUNY at Buffalo,
Buffalo, NY 14260-1500, USA}

\date{\today}

\begin{abstract} 
  We present the next-to-leading order QCD corrected rate for the
  production of a scalar Higgs boson with a pair of high $p_T$ bottom
  and anti-bottom quarks at the Tevatron and at the Large Hadron
  Collider.  Results are given for both the Standard Model and the
  Minimal Supersymmetric Standard Model.  The exclusive $b\bar{b}h$
  production rate is small in the Standard Model, but it can be
  greatly enhanced in the Minimal Supersymmetric Standard Model for
  large $\tan\beta$, making $b\bar{b}h$ an important discovery mode.
  We find that the next-to-leading order QCD results are much less
  sensitive to the renormalization and factorization scales than the
  lowest order results, but have a significant dependence on the
  choice of the renormalization scheme for the bottom quark Yukawa
  coupling.
\end{abstract}
%
\maketitle

\section{Introduction}
\label{sec:intro}

One of the most important problems of particle physics is to uncover
the origin of the electroweak symmetry breaking.  In the simplest
version of the Standard Model (SM) of particle physics, the breaking
of the electroweak symmetry introduces a single physical scalar
particle, the Higgs boson, that couples to both gauge bosons and
fermions. Extensions of the Standard Model, like the Minimal
Supersymmetric Standard Model (MSSM), introduce several scalar and
pseudoscalar Higgs bosons. Finding experimental evidence for one or
more Higgs particles is therefore a major goal of current and future
accelerators. Direct searches at LEP2 require that the SM Higgs boson
mass ($M_h$) be heavier than 114.4 GeV (at 95\%
c.l.)~\cite{Lephwg:2003}, while precision electroweak measurements
imply $M_h\!<\!219$~GeV (at 95\% c.l.)~\cite{Lepewwg:2003}. The light
scalar Higgs particle of the MSSM ($h^0$) should have mass between the
theoretical upper bound of about 130~GeV and the experimental lower
bound from LEP2, $M_{h^0}\!>\!91$~GeV (at 95\% c.l.,
$0.5\!<\tan\beta\!<\!2.4$ excluded)~\cite{Lephwg2:2001}. In both
cases, a Higgs boson should lie in a mass region which will certainly
be explored at either the Fermilab $p\bar{p}$ Tevatron collider or at
the CERN $pp$ Large Hadron Collider (LHC).

The dominant production mechanism for a SM Higgs boson in hadronic
interactions is gluon fusion. Among the subleading modes, the
associated production with either electroweak gauge bosons or top
quark pairs, as well as weak boson fusion, play crucial roles.  The
inclusion of higher order QCD corrections is in general essential to
stabilize the theoretical predictions of the corresponding rates.  All
of them have now been calculated at next-to-leading order
(NLO)~\cite{Han:1991ia,Han:1992hr,Dawson:1991zj,Djouadi:1991tk,Graudenz:1993pv,
  Spira:1995rr,Beenakker:2001rj,Reina:2001sf,Beenakker:2002nc,Reina:2001bc,
  Dawson:2002tg,Dawson:2003zu} and, in the case of gluon fusion and
associated production with gauge bosons, at next-to-next-to-leading
order
(NNLO)~\cite{Harlander:2002wh,Anastasiou:2002yz,Harlander:2001is,
  Catani:2001ic,Ravindran:2002dc,Brein:2003wg} in perturbative QCD.

If the Standard Model is not the full story, however, then other
mechanisms of Higgs production become very important. Here, we focus
on Higgs boson production with a pair of bottom quark and antiquark.
The coupling of the Higgs boson to a $b\bar{b}$ pair is suppressed in
the Standard Model by the small factor, $m_b/v$, where
$v\!=\!(\sqrt{2}G_F)^{-1/2}\!=\!246$~GeV, implying that the SM Higgs
production rate in association with bottom quarks is very small at
both the Tevatron and the LHC. In a two Higgs doublet model or in the
MSSM, however, this coupling grows with the ratio of neutral Higgs
boson vacuum expectation values, $\tan\beta$, and can be significantly
enhanced over the Standard Model coupling, leading to an observable
production rate for a Higgs boson in association with bottom quarks in
some regions of the parameter space.

The production of a Higgs boson in association with bottom quarks at
hadron colliders has been the subject of much recent theoretical
interest. At the tree level, the cross section is almost entirely
dominated by $gg\to b\bar{b}h$, with only a small contribution from
$q\bar{q}\to b\bar{b}h$, at both the Tevatron and the LHC. The
integration over the phase space of the final state bottom quarks
gives origin to large logarithms proportional to $\ln(m_b/\mu_h)$
(where $\mu_h\!\simeq\!M_h$), which arise from the splitting of an
initial gluon into a pair of almost on-shell collinear bottom quarks.
The use of bottom quark parton distribution functions in the proton
(or anti-proton) sums these large logarithms to all orders, and could
therefore improve a fixed order calculation.  The inclusive cross
section for $b\bar{b}h$ production should then be dominated by the
bottom quark fusion process $b\bar{b}\to h$, as originally proposed in
Ref.~\cite{Dicus:1989cx}.  Some important progress has been achieved
recently.  The $b\bar{b}\to h$ production process has been calculated
at NNLO in QCD~\cite{Harlander:2003ai}.  At
NLO~\cite{Dicus:1998hs,Balazs:1998sb}, the residual factorization
scale dependence is quite large, but at NNLO there is almost no scale
dependence. Interestingly enough the NNLO results show that the
perturbative cross section is better behaved when the factorization
scale is $\mu_f\!\simeq\!M_h/4$ (and the renormalization scale is
$\mu_r\!\simeq\!M_h$), as expected on quite general theoretical
grounds~\cite{Rainwater:2002hm,Plehn:2002vy,Maltoni:2003pn,Boos:2003yi}.
Moreover, the inclusive $b\bar{b}h$ cross section has been obtained at
NLO in QCD via a fixed order calculation that includes the
$O(\alpha_s)$ corrections to the parton level processes
$gg,q\bar{q}\to
b\bar{b}h$~\cite{Kraemer:2003lh,Spira:2003cn,Dittmaier:2003ej}. The
obtained results are compatible with $b\bar{b}\to h$ at NNLO, and show
that there is actually no large discrepancy between the NLO fixed
order calculation and the use of $b$-quark parton distribution
functions, contrary to what was originally claimed.  However, the
results of the fixed order calculation have a substantial scale
dependence and a better control of the residual large uncertainty is
desirable for a complete understanding of the comparison between the
two approaches.

In spite of its theoretical interest, the inclusive cross section is
experimentally relevant only if a Higgs boson can be detected above
the background without tagging any of the outgoing bottom quarks.
Higgs production from $b\bar{b}$ fusion could be useful, for instance,
in a supersymmetric model with a large value of $\tan\beta$, when
combined with the decays $h^0,H^0\to\mu^+\mu^-$ and
$h^0,H^0\to\tau^+\tau^-$~\cite{cms:1994,Barger:1998pp,
  Richter-Was:1998ak,atlas:1999}.  However, even in this case, the
inclusive measurement of a Higgs signal would not determine the bottom
quark Yukawa coupling unambiguously, since it should be interpreted as
the result of the combined action of other production channels besides
$b\bar{b}\to h^0,H^0$ (e.g. $gg\to h^0,H^0$).

Requiring one or two high $p_T$ bottom quarks in the final state
reduces the signal cross section with respect to $b\bar{b}\to h$, but
it also greatly reduces the
background~\cite{Carena:1998gk,atlas:1999}.  Moreover, it assures that
the detected Higgs boson has been radiated off a bottom or anti-bottom
quark and the corresponding cross section is therefore unambiguously
proportional to the bottom quark Yukawa coupling.  Using arguments
similar to the ones illustrated above for the case of the inclusive
cross section, one can argue that if the final state has one high
$p_T$ bottom quark then the relevant subprocess is $gb\to
bh$~\cite{Dicus:1998hs}. The cross section for $gb\to bh$ has been
computed including NLO QCD corrections~\cite{Campbell:2002zm} and the
residual uncertainty due to higher order QCD corrections is small. On
the other hand, if the final state has two high $p_T$ bottom quarks
and a Higgs boson, then no final state bottom quark can originate from
a bottom quark parton distribution function. The lowest order relevant
parton level processes are unambiguously $gg\to b\bar{b}h$ and
$q\bar{q}\to b\bar{b}h$.  While the rate for this final state is
considerably smaller than for the $b\bar{b}\to h$ and $gb\to bh$
subprocesses, the background is correspondingly reduced.  The final
states can be further categorized according to the decay of the Higgs
boson. Existing studies have considered mostly the dominant Higgs
decay channel, $h\to
b\bar{b}$~\cite{Dai:1995vu,Dai:1996rn,Carena:1998gk,Affolder:2000rg,
  Carena:2000yx,Balazs:1998nt, atlas:1999,Richter-Was:1997gi}, but
also $h\to \tau^+\tau^-$~\cite{Carena:1998gk} and $h\to
\mu^+\mu^-$~\cite{Dawson:2002cs,Boos:2003jt}.

In this paper, we present the NLO QCD corrected rates and phase space
distributions for the fully exclusive processes $pp,p\bar{p}\to
b\bar{b}h$, where the final state includes two high $p_T$ bottom
quarks. In order to reproduce as closely as possible the currently
used experimental cuts, we require the final state bottom
quark/anti-quark to have a transverse momentum higher than
$p_T^{cut}\!=\!20$~GeV and a pseudorapidity $|\eta|\!\le\!2$ for the
Tevatron and $|\eta|\!\le\!2.5$ for the LHC.  The cut on
$p_T^{b,\bar{b}}$ greatly affects the cross section and we therefore
study the dependence of the cross section on this cut.  Similar
results have been recently presented in Ref.~\cite{Dittmaier:2003ej},
where however no cut on the pseudorapidity has been imposed.  Our
discussion will focus on assessing the uncertainty of the theoretical
prediction for the exclusive $pp,p\bar{p}\to b\bar{b}h$ rates, after
the full set of NLO QCD corrections has been included.  We will show
how the large dependence on the unphysical renormalization and
factorization scales present in the lowest order (LO) calculation of
the cross section is greatly reduced at NLO.  Moreover, we will study
the dependence on the choice of renormalization scheme for the bottom
quark Yukawa coupling.  While for Higgs decays and Higgs production in
$e^+e^-$ collisions using the $\overline{MS}$ definition of the bottom
quark Yukawa coupling is an efficient way of improving the
perturbative calculation of the corresponding rate by resumming large
logarithms at all orders~\cite{Braaten:1980yq,Drees:1990dq,
  Gorishnii:1984cu,Kataev:1994be}, this may be less compelling in the
case of hadronic Higgs production.  Finally, we will extend our
calculation to the scalar sector of the MSSM, including the SM QCD
corrections at NLO.  Preliminary results of the study described in
this paper have been already presented at several
conferences~\cite{talksbbh:2003}.

The plan of the paper is as follows.  In Section~\ref{sec:calculation}
we present an overview of our calculation. Since the NLO QCD
corrections to $q\bar{q},gg\to b\bar{b}h$ proceed in strict analogy to
those for $q\bar{q}\to
t\bar{t}h$~\cite{Beenakker:2001rj,Reina:2001sf,Reina:2001bc,Beenakker:2002nc}
and $gg\to
t\bar{t}h$~\cite{Beenakker:2001rj,Beenakker:2002nc,Dawson:2002tg,Dawson:2003zu},
we will be very brief on details and devote more time to the
discussion of the residual theoretical uncertainty, emphasizing those
aspects that are characteristic of the $b\bar{b}h$ production process.
Numerical results for the Tevatron and the LHC will be presented in
Section~\ref{sec:results}, for both the SM Higgs boson and the scalar
MSSM Higgs bosons in some prototype regions of the model parameter
space. Section~\ref{sec:conclusions} contains our conclusions.

\section{Calculation}
\label{sec:calculation}

\subsection{Basics}
\label{subsec:basics}

The total cross section for $pp,p\bar{p}\to b\bar{b}h$ at ${\cal
O}(\alpha_s^3)$ can be written as:
\begin{eqnarray}
\label{eq:sigma_nlo}
&&\sigma_{\sss NLO}(pp,p\bar{p}\to b\bar{b}h)=\nonumber\\
&&\sum_{ij}\frac{1}{1+\delta_{ij}}
\int dx_1 dx_2 \left[{\cal F}_i^p(x_1,\mu) {\cal F}_j^{p,\bar{p}}(x_2,\mu)
{\hat \sigma}^{ij}_{\sss NLO}(x_1,x_2,\mu)+(1\leftrightarrow 2)\right]
\,\,\,,
\end{eqnarray}
where ${\cal F}_i^{p,\bar{p}}$ are the NLO parton distribution
functions (PDFs) for parton $i$ in a proton or anti-proton, defined at
a generic factorization scale $\mu_f\!=\!\mu$, and ${\hat
  \sigma}^{ij}_{\sss NLO}$ is the ${\cal O}(\alpha_s^3)$ parton-level
total cross section for incoming partons $i$ and $j$, made of the
channels $q\bar{q},gg\to b\bar{b}h$ and $(q,\bar{q})g\to
b\bar{b}h(q,\bar{q})$, and renormalized at an arbitrary scale $\mu_r$
which we also take to be $\mu_r\!=\!\mu$.  Throughout this paper we
will always assume the factorization and renormalization scales to be
equal, $\mu_r\!=\!\mu_f\!=\!\mu$.  The partonic center-of-mass energy
squared, $s$, is given in terms of the hadronic center-of-mass energy
squared, $s_{\sss H}$, by $s=x_1 x_2 s_{\sss H}$. At both the Tevatron
and the LHC, the dominant contribution is from the gluon-gluon initial
state, although we include all initial states.

The NLO parton-level total cross section reads
\begin{equation}
\label{eq:sigmahat_nlo}
{\hat\sigma}_{\sss NLO}^{ij}(x_1,x_2,\mu)=
{\hat \sigma}_{\sss LO}^{ij}(x_1,x_2,\mu)+
\delta {\hat \sigma}_{\sss NLO}^{ij}(x_1,x_2,\mu)\,\,\,,
\end{equation}
where ${\hat\sigma}_{\sss LO}^{ij}( x_1,x_2,\mu)$ is the ${\cal
  O}(\alpha_s^2)$ Born cross section, and $\delta{\hat\sigma}_{\sss
  NLO}^{ij}(x_1,x_2,\mu)$ consists of the ${\cal O}(\alpha_s)$
corrections to the Born cross sections for $gg,q\bar{q}\to b\bar{b}h$
and of the tree level $(q,\bar{q})g\to b\bar{b}h(q,\bar{q})$
processes, including the effects of mass factorization.

\begin{figure}[t]
\begin{center}
\includegraphics[scale=0.8]{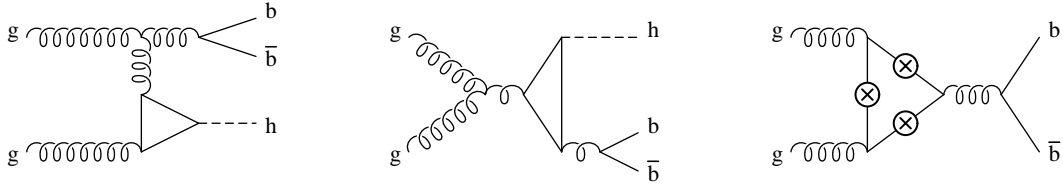} 
\caption[]{Sample of diagrams corresponding to ${\cal O}(\alpha_s)$ 
  virtual corrections where the Higgs boson couples to an internal
  fermion loop and not to the external $b\bar{b}$ pair. The circled
  cross denotes all possible insertion of the final state Higgs boson
  leg, each insertion corresponding to a different diagram. }
\label{fg:bbh_fermion_loop}
\end{center}
\end{figure}

The evaluation of ${\hat\sigma}^{ij}_{\sss NLO}$ proceeds along the
same lines as the corresponding calculation for $t\bar{t}h$
production~\cite{Beenakker:2001rj,Reina:2001sf,Beenakker:2002nc,
  Reina:2001bc,Dawson:2002tg,Dawson:2003zu} and we refer to
Refs.~\cite{Reina:2001bc,Dawson:2003zu} for a detailed description of
the techniques used in our calculation. We notice that, in view of the
generalization to the MSSM with a very enhanced bottom quark Yukawa
coupling, both top and bottom quark loops need to be considered in
those virtual diagrams where the Higgs boson couples directly to a
closed loop of fermions, a sample of which is illustrated in
Fig.~\ref{fg:bbh_fermion_loop}.

Contrary to the case of $t\bar{t}h$ production, the NLO cross section
for $b\bar{b}h$ production depends significantly on the
renormalization scheme used for the bottom quark Yukawa coupling, i.e.
for the bottom quark mass appearing in $g_{b\bar{b}h}\!=\!m_b/v$.  In
our calculation of the NLO $p\bar{p},pp\to b\bar{b}h$ cross section we
have considered, for the renormalization of the bottom quark Yukawa
coupling, both the on-shell and the $\overline{MS}$ subtraction
schemes (in the $t\bar{t}h$ case we only used the on-shell top quark
renormalized mass everywhere~\cite{Dawson:2003zu}). The
$\overline{MS}$ scheme results in a running bottom quark Yukawa
coupling and potentially gives better control over higher order
contributions beyond the 1-loop corrections.  We will study the origin
and magnitude of the residual scheme dependence in
Sections~\ref{subsec:ren_scheme} and \ref{subsec:results_sm}.

\subsection{Renormalization scheme dependence}
\label{subsec:ren_scheme}

The ultraviolet (UV) divergences arising from self-energy and vertex
${\cal O}(\alpha_s)$ virtual corrections to $q\bar{q},gg\to b\bar{b}h$
are regularized in $d\!=4-2\epsilon$ dimensions and renormalized by
introducing counterterms for the wave functions of the external fields
($\delta Z_2^{(q)}$ (for $q\!=\!u,d,c,s$), $\delta Z_2^{(b)}$ for the
bottom quark, and $\delta Z_3$ for the gluon), for the bottom quark
mass, $\delta m_b$, and for the bottom quark Yukawa and strong
coupling constants, $\delta g_{b\bar{b}h}$ and $\delta Z_{\alpha_s}$.
We follow the same renormalization prescription and notation adopted
in Refs.~\cite{Reina:2001bc,Dawson:2003zu} for the NLO $t\bar{t}h$
inclusive cross section. Consequently, we fix the wave-function
renormalization constants of the external massless quark fields,
$\delta Z_2^{(q)}$, using on-shell subtraction, while we define the
wave function renormalization constant of external gluons, $\delta
Z_3$, using the $\overline{MS}$ subtraction scheme and the $\alpha_s$
renormalization constant, $\delta Z_{\alpha_s}$, using the
$\overline{MS}$ scheme modified to decouple the top
quark~\cite{Collins:1978wz,Nason:1989zy}.  Explicit expressions for
$\delta Z_2^{(q)}$, $\delta Z_3$, and $\delta Z_{\alpha_s}$ can be
found in Refs.~\cite{Reina:2001bc,Dawson:2003zu}.

However, given the large sensitivity of the $\overline{MS}$ bottom
quark mass to the renormalization scale and given the prominent role
it plays in the $b\bar{b}h$ production cross section through the
overall bottom quark Yukawa coupling, we investigate here the
dependence of the final results on the renormalization prescription
adopted for the bottom quark. We consider both the on-shell ($OS$) and
$\overline{MS}$ subtraction schemes, for both the bottom quark mass
and wave function renormalization constants.

When using the $OS$ subtraction scheme, we fix the wave function
renormalization constant of the external bottom quark field, $(\delta
Z_2^{(b)})_{OS}$, and the mass renormalization constant, $(\delta
m_b)_{OS}$, by requiring that
\begin{equation}
\label{eq:os_condition}
\hat{\Sigma}_b(\slashed{p}=m_b)=0 \;\;\; ; \;\;\; 
\lim_{\slashed{p}\to m_b}
\frac{\hat{\Sigma}_b(\slashed{p})}{\slashed{p}-m_b}=0\,\,\,,
\end{equation}
where
\begin{equation}
\label{eq:bself}
\hat{\Sigma}_b=(\slashed{p}-m_b) 
\left(\Sigma_V+\delta Z_2^{(b)}\right)+m_b \left(
\Sigma_S+\Sigma_V-\frac{\delta m_b}{m_b}\right)
\end{equation}
denotes the renormalized bottom quark self-energy at 1-loop in QCD,
expressed in terms of the vector, $\Sigma_V$, and scalar, $\Sigma_S$,
parts of the unrenormalized self-energy, and of the mass and wave
function renormalization constants. Using Eq.~(\ref{eq:os_condition})
in $d\!=\!4-2\epsilon$ dimensions one finds
\begin{eqnarray}
\label{eq:dz2b_os}
\left(\delta Z_{2}^{(b)}\right)_{OS}&=&
-\frac{\alpha_s}{4\pi}\,C_F\left(\frac{4\pi\mu^2}{m_b^2}\right)^{\epsilon}
\Gamma(1+\epsilon) 
 \left(\frac{1}{\epsilon_{\sss UV}}+4+\frac{2}{\epsilon_{\sss IR}}\right)
\,\,\,,\\ 
\label{eq:dmb_os}
\left(\frac{\delta m_b}{m_b}\right)_{OS}&=&
-\frac{\alpha_s}{4\pi}\,C_F\left(\frac{4\pi\mu^2}{m_b^2}\right)^{\epsilon}
\Gamma(1+\epsilon) 
\left(\frac{3}{\epsilon_{\sss UV}}+4\right)\,\,\,,
\end{eqnarray}
where we have explicitly distinguished between ultraviolet and infrared
divergences. The infrared divergences are cancelled between virtual
and real soft and collinear contributions according to the pattern
outlined in Refs.~\cite{Reina:2001bc,Dawson:2003zu}, to which we refer
for more details.

In the $\overline{MS}$ scheme, the bottom quark renormalization
constants are fixed by requiring that they cancel the UV divergent
parts of the bottom quark self energy $\hat{\Sigma}_b$ of
Eq.~(\ref{eq:bself}), i.e.
\begin{eqnarray}
\label{eq:dz2b_msbar}
\left(\delta Z_{2}^{(b)}\right)_{\overline{MS}}&=&
-\frac{\alpha_s}{4\pi}\,C_F\left(4\pi\right)^{\epsilon}
\Gamma(1+\epsilon) \frac{1}{\epsilon_{\sss UV}} \,\,\,,
\\
\label{eq:dmb_msbar}
\left(\frac{\delta m_b}{m_b}\right)_{\overline{MS}}&=&
-\frac{\alpha_s}{4\pi}\,C_F\left(4\pi\right)^{\epsilon}
\Gamma(1+\epsilon) \frac{3}{\epsilon_{\sss UV}} \,\,\,.
\end{eqnarray}
According to the LSZ prescription~\cite{Lehmann:1955rq}, one also
needs to consider the insertion of the renormalized one-loop
self-energy corrections on the external bottom quark legs. While these
terms are zero in the $OS$ scheme (see Eq.~(\ref{eq:os_condition})),
they are not zero in the $\overline{MS}$ scheme. Together with
$(\delta Z_{2}^{(b)})_{\overline{MS}}$, their contribution to the NLO
cross section equals the contribution of the wave function counterterm
in the $OS$ scheme, $(\delta Z_{2}^{(b)})_{OS}$, as expected from the
LSZ prescription itself. The cross section does not depend on the
renormalization of the external particle wave functions.

We therefore focus on the scheme dependence induced by the choice of
different subtraction schemes for the bottom quark mass. We note that
the bottom quark mass counterterm has to be used twice: once to
renormalize the bottom quark mass appearing in internal propagators
and once to renormalize the bottom quark Yukawa coupling. Indeed, if
one considers only QCD corrections, the counterterm for the bottom
quark Yukawa coupling,
\begin{equation}
\label{eq:b_yuk_sm}
\delta g_{b\bar{b}h}\!=\!\frac{\delta m_b}{v}\,\,\,,
\end{equation}
coincides with the counterterm for the bottom quark mass, since the SM
Higgs vacuum expectation value $v$ is not renormalized at 1-loop in
QCD. This stays true when we generalize the $g_{b\bar{b}h}$ coupling
from the SM to the case of the scalar Higgs bosons of the MSSM.

At 1-loop order in QCD, the relation between the pole mass, $m_b$, and
the $\overline{MS}$ mass, $\overline{m}_b(\mu)$, is indeed determined
by the difference between the $OS$ and $\overline{MS}$ bottom mass
counterterms,
$\frac{\displaystyle{\alpha_s}}{\displaystyle{4\pi}}\delta CT$, since
\begin{equation}
\label{eq:mb_msbar}
\overline{m}_b(\mu)=m_b \left\{1-\frac{\alpha_s(\mu)}{4 \pi} C_F
\left[3\ln\left(\frac{\mu^2}{m_b^2}\right)+4\right]\right\}
\equiv m_b \left[1-\frac{\alpha_s(\mu)}{4\pi} \delta CT(\mu)\right]\,\,\,.
\end{equation}
Adopting the $OS$ or $\overline{MS}$ prescription consists of using
either Eq.~(\ref{eq:dmb_os}) or Eq.~(\ref{eq:dmb_msbar}) for the
bottom mass counterterms while substituting $m_b$ or
$\overline{m}_b(\mu)$ respectively in both the bottom quark propagator
and Yukawa coupling.  At ${\cal O}(\alpha_s^3)$ the two prescriptions
give identical results.  Indeed, replacing $m_b$ by
$\overline{m}_b(\mu)$ in the Yukawa coupling adds a term
\begin{equation}
\label{eq:msbar_yuk_as3}
-\frac{\alpha_s(\mu)}{2\pi}\delta CT(\mu)\hat{\sigma}_{\sss LO}+
{\cal O}(\alpha_s^4)
\end{equation}
to the NLO parton level cross section, which compensates exactly for
the difference in the OS and $\overline{MS}$ counterterms. On the
other hand, using the $\overline{MS}$ mass in the bottom quark
propagator,
\begin{equation}
\label{eq:msbar_prop_as3}
\frac{i}{\slashed{p}-\overline{m}_b(\mu)} = \frac{i}{\slashed{p}-m_b} 
\left[1+i m_b \frac{\alpha_s}{4\pi} \delta CT(\mu) 
\frac{i}{\slashed{p}-m_b}\right]+{\cal O}(\alpha_s^2)\,\,\,,
\end{equation}
of the LO cross section leads to an extra contribution to the
$\overline{MS}$ NLO cross section which, together with the
$\overline{MS}$ mass counterterm insertions into the internal bottom
quark propagators (see diagrams $S_1$ in Fig.~2 of
Ref.~\cite{Reina:2001bc} and $S_2$, $S_3$, and $S_4$ in Fig.~2 of
Ref.~\cite{Dawson:2003zu}), coincides with the corresponding mass
counterterm insertions in the $OS$ scheme at ${\cal O}(\alpha_s^3)$.

Therefore, using OS or $\overline{MS}$ at ${\cal O}(\alpha_s^3)$ is
perturbatively consistent, the difference between the two schemes
being of higher order and hence, strictly speaking, part of the
theoretical uncertainty of the NLO calculation. One notices however
that some of the large logarithms involved in the renormalization
procedure of the NLO cross section come from the renormalization of
the bottom quark mass, and are nicely factored out by using the
$\overline{MS}$ bottom mass in the bottom quark Yukawa coupling (see
Eq.~(\ref{eq:mb_msbar})).  Therefore one should consider reorganizing
the perturbative expansion in terms of leading logarithms (of the form
$\alpha_s^n(\mu)\ln^n(\mu^2/m_b^2)$) or next-to-leading-logarithms (of
the form $\alpha_s^n(\mu)\ln^{n-1}(\mu^2/m_b^2)$, for
$\mu\!\simeq\!M_h$), as obtained by replacing the $\overline{MS}$
bottom mass in the Yukawa coupling by the corresponding 1-loop or
2-loop renormalization group improved $\overline{MS}$ masses:
\begin{eqnarray}
\label{eq:mb_msbar_rg_1l}
\overline{m}_b(\mu)_{1l}&=& m_b 
\left[\frac{\alpha_s(\mu)}{\alpha_s(m_b)}\right]^{c_0/b_0}\,\,\,,\\
\label{eq:mb_msbar_rg_2l}
\overline{m}_b(\mu)_{2l}&=& m_b
\left[\frac{\alpha_s(\mu)}{\alpha_s(m_b)}\right]^{c_0/b_0}
\left[ 1+\frac{c_0}{b_0}\left(c_1-b_1\right)
\frac{\alpha_s(\mu)-\alpha_s(m_b)}{\pi}\right]
\left(1-\frac{4}{3}\frac{\alpha_s(m_b)}{\pi}\right)\,\,\,,
\nonumber\\
\end{eqnarray}
where 
\begin{eqnarray}
\label{eq:mb_msbar_rg_coeff}
b_0&=&\frac{1}{4\pi}\left(\frac{11}{3}N-\frac{2}{3}n_{lf}\right)
\,\,\,,\,\,\, c_0=\frac{1}{\pi}\,\,\,,\\
b_1&=&\frac{1}{2\pi}\frac{51 N-19 n_{lf}}{11 N-2 n_{lf}}
\,\,\,,\,\,\, c_1=\frac{1}{72\pi}\left(101 N-10 n_{lf}\right)\,\,\,,
\end{eqnarray}
are the one and two loop coefficients of the QCD $\beta$-function and
mass anomalous dimension $\gamma_m$, while $N\!=\!3$ is the number of
colors and $n_{lf}\!=\!5$ is the number of light flavors.

In both Higgs boson decays to heavy quarks and Higgs boson production
with heavy quarks in $e^+e^-$ collisions, using
Eq.~(\ref{eq:mb_msbar_rg_1l}) at LO and Eq.~(\ref{eq:mb_msbar_rg_2l})
at NLO in the Yukawa coupling proves to be a very powerful way to
stabilize the perturbative calculation of the cross
section~\cite{Braaten:1980yq}. The difference between LO and NLO rates
is reduced and the dependence on the renormalization and factorization
scales at NLO is very mild, indicating a very small residual
theoretical error or equivalently a very good convergence of the
perturbative expansion of the corresponding rate.  This is due to the
fact that in these cases to a large extent the ${\cal O}(\alpha_s)$
QCD corrections amount to a renormalization of the heavy quark mass in
the Yukawa coupling. In more complicated cases, like the case of the
hadronic cross section discussed in this paper, the previous argument
is not automatically true.

\begin{figure}[t]
\begin{center}
\includegraphics[scale=0.6]{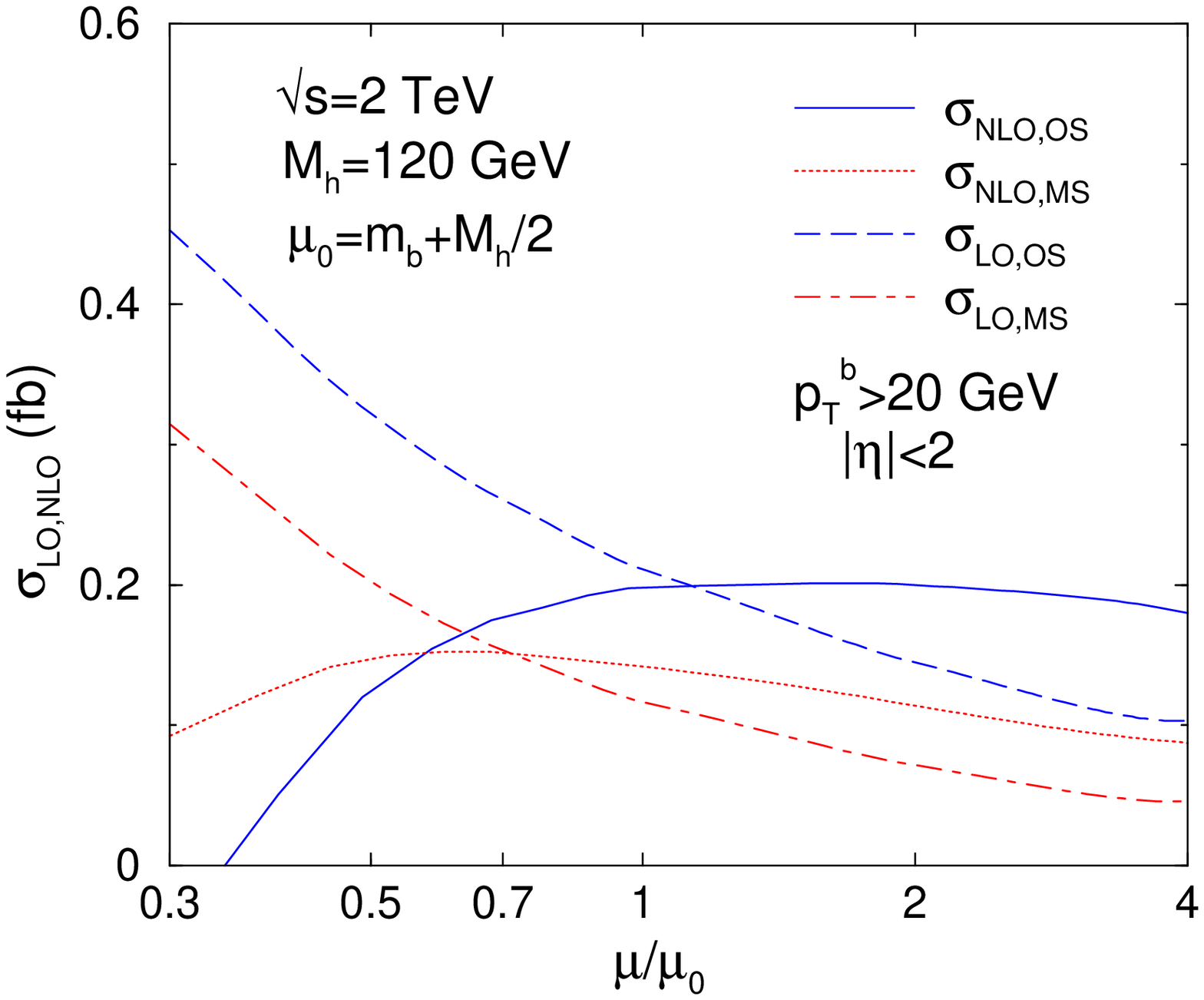} 
\includegraphics[scale=0.6]{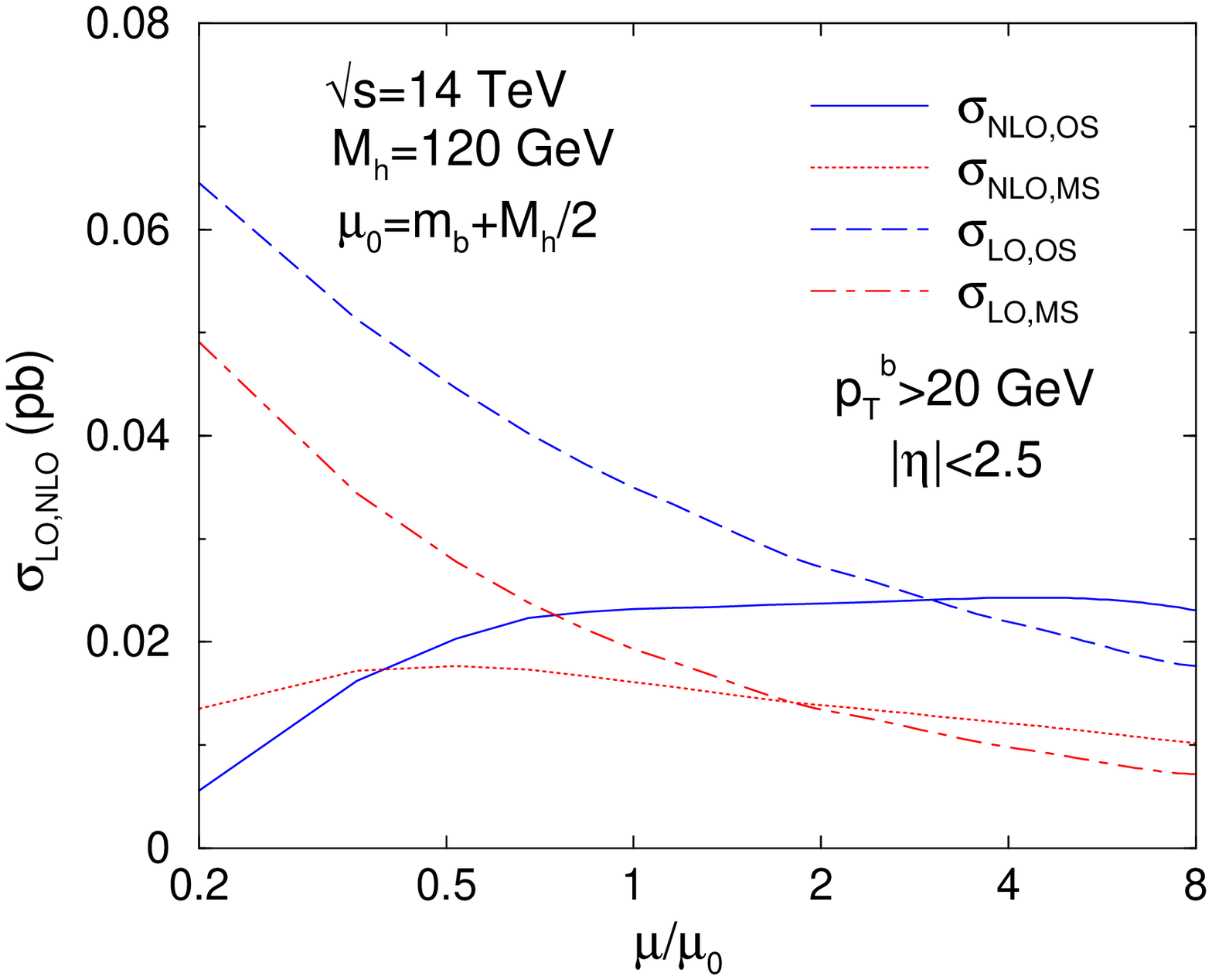} 
\caption[]{$\sigma_{\sss NLO}$ and $\sigma_{\sss LO}$ for $p\bar{p}\to
  b\bar{b}h$ at $\sqrt{s}\!=\!2$~TeV (top) and for $pp\to b\bar{b}h$
  at $\sqrt{s}\!=\!14$~TeV (bottom) as a function of the
  renormalization/factorization scale $\mu$, for $M_h=120$~GeV.  The
  curves labeled $\sigma_{\sss LO,OS}$ and $\sigma_{\sss NLO,OS}$ use
  the $OS$ renormalization scheme for the bottom quark Yukawa coupling,
  while the curves labeled $\sigma_{\sss LO,MS}$ and $\sigma_{\sss
    NLO,MS}$ use the $\overline{MS}$ scheme.}
\label{fg:bbh_mu_dep}
\end{center}
\end{figure}

Using the $OS$ or $\overline{MS}$ bottom quark mass mainly affects the
Yukawa coupling.  Therefore, in the hadronic case, we will look at the
different behavior of the NLO cross section when the bottom quark
Yukawa coupling is renormalized either in the $OS$ or in the
$\overline{MS}$ scheme, keeping the bottom quark pole mass everywhere
else.  Fig.~\ref{fg:bbh_mu_dep} of Section~\ref{sec:results} shows the
renormalization and factorization scale dependence of the LO and NLO
cross sections for $pp,p\bar{p}\to b\bar{b}h$ obtained using in the
Yukawa coupling either the pole mass $m_b$ or the $\overline{MS}$
running mass $\overline{m}_b(\mu)$ in Eq.~(\ref{eq:mb_msbar_rg_1l})
(at LO) and (\ref{eq:mb_msbar_rg_2l}) (at NLO).  The use of
$\overline{m}_b(\mu)$ both at LO and NLO seems to improve the
perturbative calculation of the cross section, since the NLO
$\overline{MS}$ cross section is better behaved than the NLO $OS$
cross section at low scales and since the difference between LO and
NLO cross section is smaller when the bottom quark Yukawa coupling is
renormalized in the $\overline{MS}$ scheme than in the $OS$ scheme.
However, both the OS and the $\overline{MS}$ cross sections have very
well defined regions of minimum sensitivity to the variation of the
renormalization/factorization scale and these regions do not quite
overlap. The difference between the $OS$ and $\overline{MS}$ results
at the plateau should rather be interpreted, in the absence of a NNLO
calculation, as an upper bound on the theoretical uncertainty.

\begin{figure}[t]
\begin{center}
\includegraphics[scale=0.6]{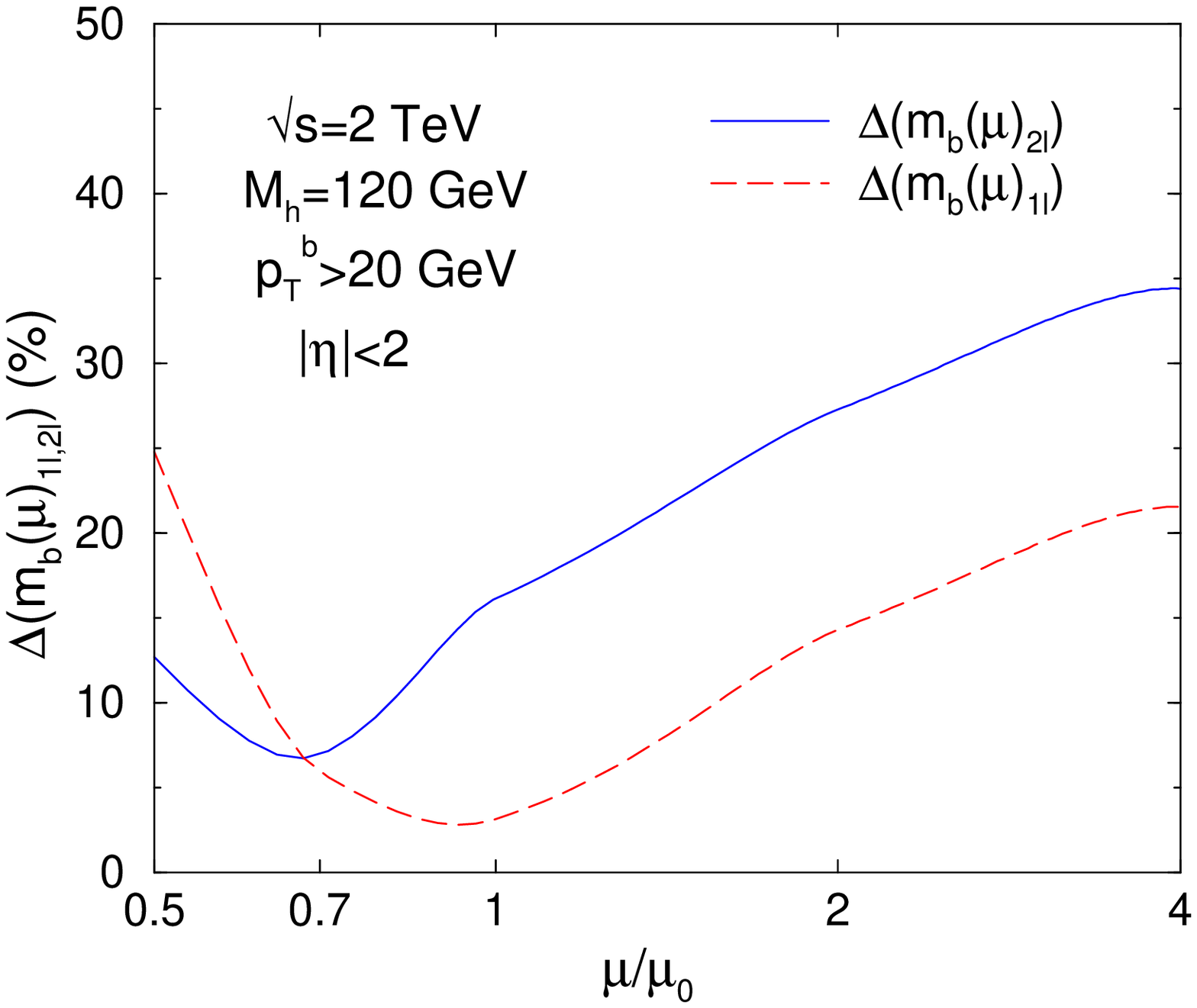} 
\includegraphics[scale=0.6]{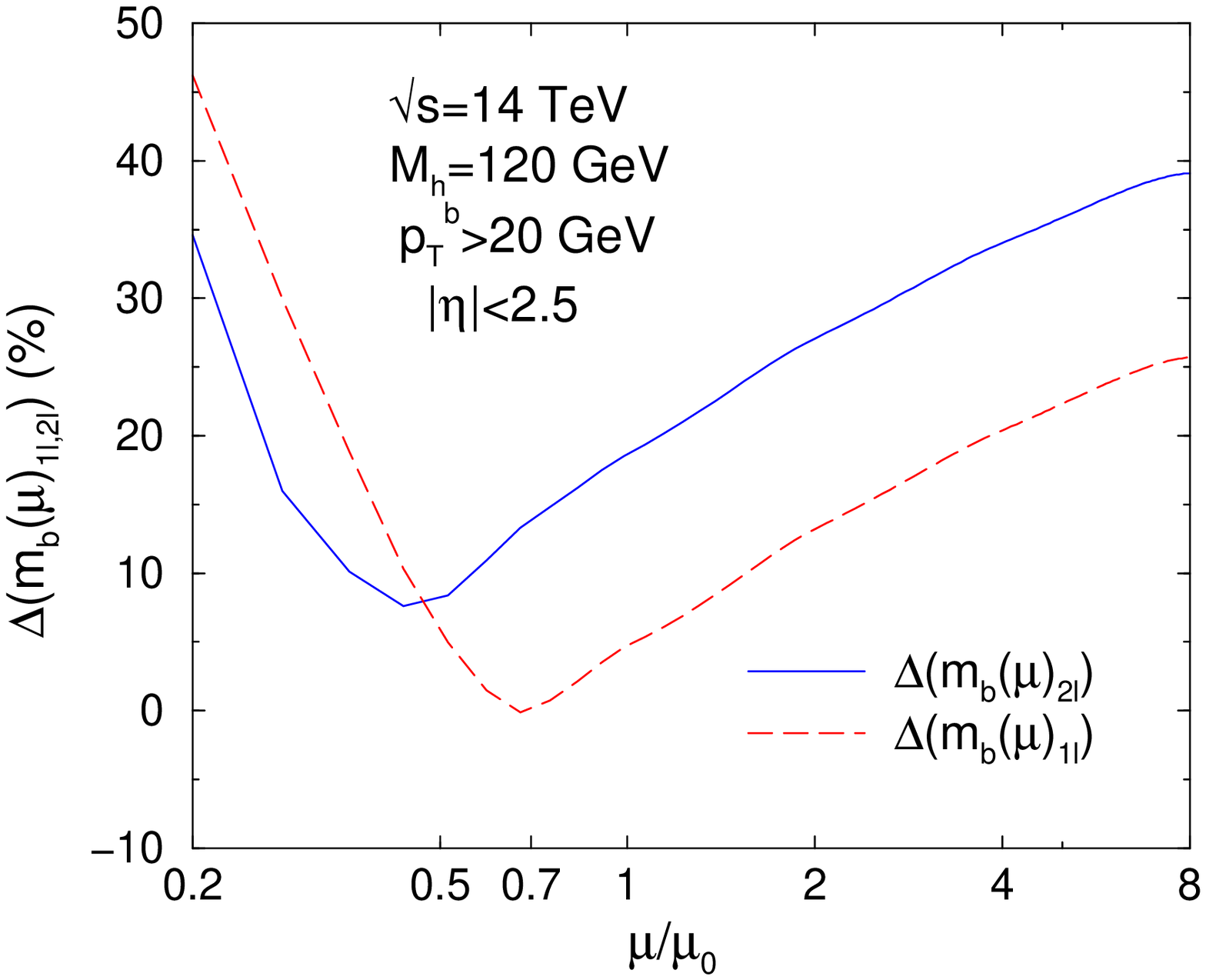} 
\caption[]{ 
  The absolute value of the percentage difference $\Delta(\%)\!=\!(\sigma_{\sss
    NLO,OS}-\sigma_{\sss NLO,\overline{MS}})/(\sigma_{\sss
    NLO,OS}+\sigma_{\sss NLO,\overline{MS}})$ for $p\bar{p}\to
  b\bar{b}h$ at $\sqrt{s}=2$~TeV (top) and for $pp\to b\bar{b}h$ at
  $\sqrt{s}=14$~TeV (bottom) as a function of the
  renormalization/factorization scale $\mu$, for $M_h=120$~GeV. The
  $OS$ and $\overline{MS}$ labels refer to the renormalization scheme
  chosen for the bottom quark Yukawa coupling.  The curves labeled as
  $\Delta(m_b(\mu)_{1l})$ and $\Delta(m_b(\mu)_{2l})$ use the
  $\overline{MS}$ bottom quark Yukawa coupling with the 1-loop running mass
  of Eq.~(\ref{eq:mb_msbar_rg_1l}) and the 2-loop running mass of
  Eq.~(\ref{eq:mb_msbar_rg_2l}), respectively.}
\label{fg:bbh_delta_mu_dep}
\end{center}
\end{figure}

The origin of the large difference between the $OS$ and
$\overline{MS}$ NLO cross sections illustrated in
Fig.~\ref{fg:bbh_mu_dep} can be understood by studying the numerical
effect of the higher order terms that are included in the NLO
$\overline{MS}$ cross section when $\overline{m}_b(\mu)$ is used in
the Yukawa coupling. The parton level NLO cross sections for $ij \to
b\bar bh$ ($ij\!=\!q\bar{q},gg$) in the $OS$ and $\overline{MS}$
prescription explained above can be written as:
\begin{eqnarray}
\label{eq:sigma_os}
{\hat\sigma}^{ij}_{\sss NLO,OS}(x_1,x_2,\mu)&=&
m_b^2 \alpha_s^2(\mu)\left\{\phantom{\frac{1}{2}}\, 
g^{ij}_{\sss LO}(x_1,x_2) \right.\nonumber \\ 
&+& \left.
 \frac{\alpha_s(\mu)}{4\pi}\left[g^{ij}_{\sss NLO}(x_1,x_2,\mu)
-2 g^{ij}_{\sss LO}(x_1,x_2)\delta CT(\mu)+
\frac{m_t}{m_b} g^{ij}_{\sss cl}(x_1,x_2) \right] \right\}\,\,\,,
\nonumber \\
\\ 
\label{eq:sigma_ms}
{\hat\sigma}^{ij}_{\sss NLO,\overline{MS}}(x_1,x_2,\mu)&=&
\overline{m}_b^2(\mu) \alpha_s^2(\mu)\left\{\phantom{\frac{1}{2}}
\,g^{ij}_{\sss LO}(x_1,x_2)\right. \nonumber \\
&+& \left. 
\frac{\alpha_s(\mu)}{4\pi}\left[g^{ij}_{\sss NLO}(x_1,x_2,\mu)+ 
\frac{m_t}{\overline{m}_b(\mu)} g^{ij}_{\sss cl}(x_1,x_2)
\right]\right\}\,\,\,,
\end{eqnarray}
where the dependence on the renormalization scale is explicitly given.
$\alpha_s(\mu)$ is the 2-loop strong coupling, $m_b$ is the bottom
pole mass, and $\overline{m}_b(\mu)$ is the bottom quark
$\overline{MS}$ mass.  $g^{ij}_{\sss LO}$, $g^{ij}_{\sss NLO}$ and
$g_{\sss cl}^{ij}$ have been defined in such a way that they are the
same in the $OS$ and the $\overline{MS}$ schemes.  They correspond
respectively to the ${\cal O}(\alpha_s^2)$ ($g^{ij}_{\sss LO}$) and
${\cal O}(\alpha_s^3)$ ($g^{ij}_{\sss NLO}$) contributions to the NLO
QCD cross section, from which we have singled out the ${\cal
  O}(\alpha_s)$ virtual corrections where the Higgs boson couples to a
top quark in a closed fermion loop ($g^{ij}_{cl}$, see, e.g., diagrams
in Fig.~\ref{fg:bbh_fermion_loop}) as well as $\delta CT(\mu)$, i.e.
the difference between the $OS$ and $\overline{MS}$ bottom mass
counterterms defined in Eq.~(\ref{eq:mb_msbar}).  Using
Eqs.~(\ref{eq:sigma_os}) and (\ref{eq:sigma_ms}), one can easily
verify that the difference between the parton level NLO cross sections
obtained by using either the $OS$ or the $\overline{MS}$ scheme for
the bottom quark Yukawa coupling is, as expected, of higher order in
$\alpha_s$, i.e.:
\begin{eqnarray}
\label{eq:diff_sigma_os_msbar}
\hat\Delta&=&{\hat\sigma}^{ij}_{\sss NLO,OS}-
{\hat \sigma}^{ij}_{\sss NLO,\overline{MS}}
\nonumber \\
&=&
\alpha_s^2(\mu) g^{ij}_{\sss LO}(x_1,x_2) 
\left[m_b^2 -\overline{m}_b^2(\mu)-
m_b^2\frac{\alpha_s(\mu)}{2\pi}\delta CT(\mu)\right]
\nonumber \\
&+&\frac{\alpha_s^3(\mu)}{4\pi} (m_b^2-\overline{m}_b^2(\mu)) 
\left[ g^{ij}_{\sss NLO}(x_1,x_2,\mu)+\frac{m_t}{m_b+\overline{m}_b(\mu)}
g^{ij}_{\sss cl}(x_1,x_2)\right]\,\,\,.
\end{eqnarray}
The term in the first bracket of Eq.~(\ref{eq:diff_sigma_os_msbar})
vanishes at ${\cal O}(\alpha_s^3)$, as can be easily verified by using
Eq.~(\ref{eq:mb_msbar}).  Hence all the terms in
Eq.~(\ref{eq:diff_sigma_os_msbar}) only contribute at $O(\alpha_s^4)$
and higher.  However, while the first term is in general quite small,
the term proportional to $g_{\sss NLO}^{ij}(x_1,x_2,\mu)$ can be large
and has a non trivial scale dependence that we can formally write as:
\begin{equation}
\label{eq:f_ij_nlo}
g_{\sss NLO}^{ij}(x_1,x_2,\mu)=g_1^{ij}(x_1,x_2)+
\tilde{g}_1^{ij}(x_1,x_2)\ln\left(\frac{\mu^2}{s}\right)\,\,\,.
\end{equation}
From renormalization group arguments~\cite{Reina:2001bc,Dawson:2003zu}
one can see that $\tilde{g}_1^{ij}(x_1,x_2)$ is given by:
\begin{eqnarray}
\label{eq:f_ij_nlo_mudep}
\tilde{g}_{1}^{ij}(x_1,x_2)&=&
2\,\left\{(4\pi b_0+4) g^{ij}_{\sss LO}(x_1,x_2)-
\sum_k\left[\int_\rho^1 dz_1 P_{ik}(z_1)g^{kj}_{\sss LO}(x_1 z_1,x_2)
\right. \right.\nonumber\\
&+&\left.\left.\int_\rho^1 dz_2 P_{jk}(z_2)g^{ik}_{\sss LO}(x_1,x_2z_2) 
\right]\right\}
\,\,\,,
\end{eqnarray}
where $\rho\!=\!(2m_b+M_h)^2/s$, $P_{ij}(z)$ denotes the lowest-order
regulated Altarelli-Parisi splitting function~\cite{Altarelli:1977zs}
of parton $i$ into parton $j$, when $j$ carries a fraction $z$ of the
momentum of parton $i$, (see e.g. Section V of
Ref.~\cite{Dawson:2003zu}), and $b_0$ is given in
Eq.~(\ref{eq:mb_msbar_rg_coeff}).  As a result, $\hat\Delta$, defined
in Eq.~(\ref{eq:diff_sigma_os_msbar}), turns out to have a non trivial
scale dependence and, thus, the difference between the NLO hadronic
cross section calculated with the $OS$ or with the $\overline{MS}$
definition of the bottom quark Yukawa coupling can be numerically
quite significant for some values of the renormalization/factorization
scale, as we will illustrate in Section~\ref{sec:results} (see
Figs.~\ref{fg:bbh_mu_dep} and \ref{fg:bbh_delta_mu_dep}).

\section{Numerical results}
\label{sec:results}

\begin{figure}[t]
\begin{center}
\includegraphics[scale=0.6]{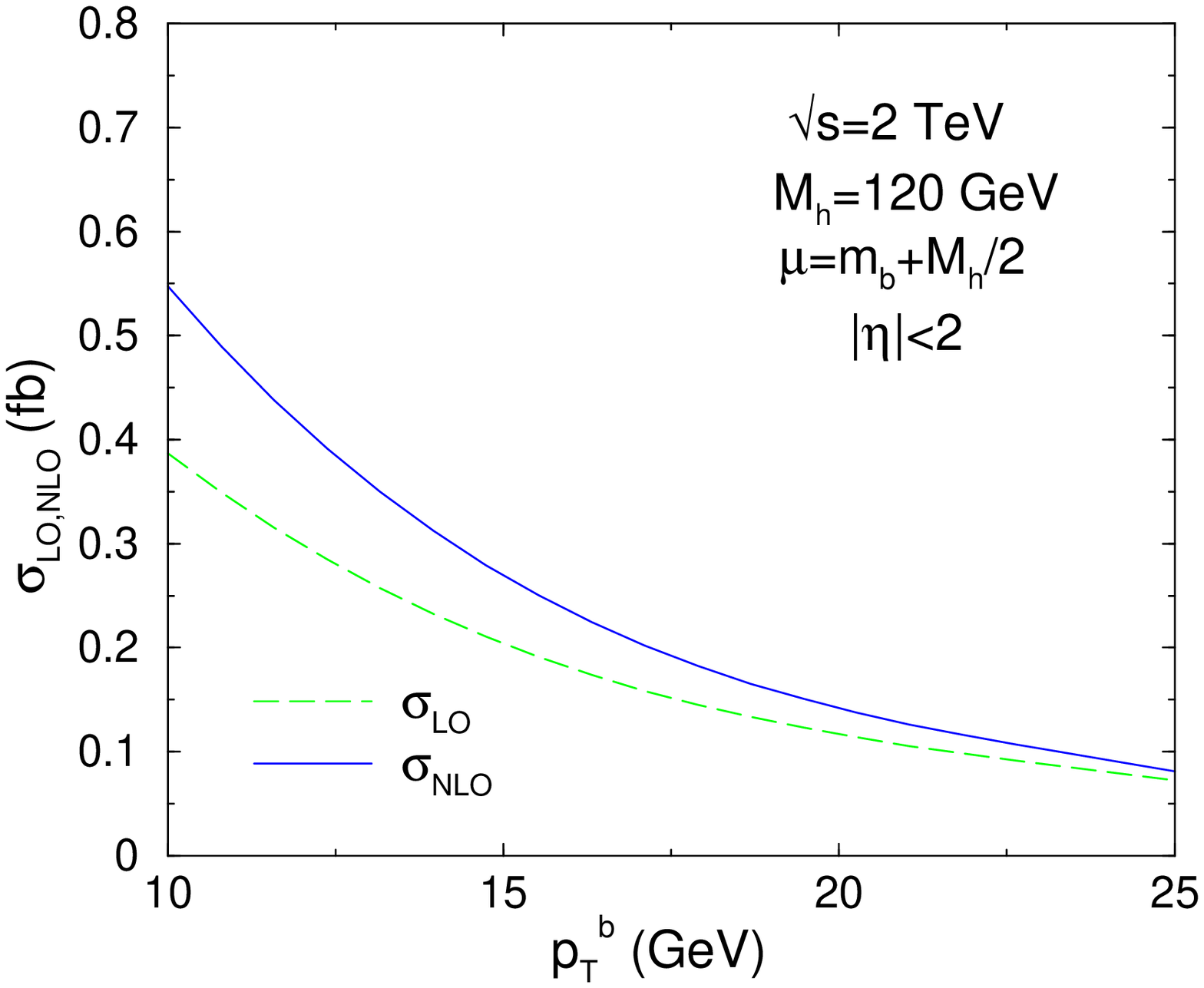} 
\includegraphics[scale=0.6]{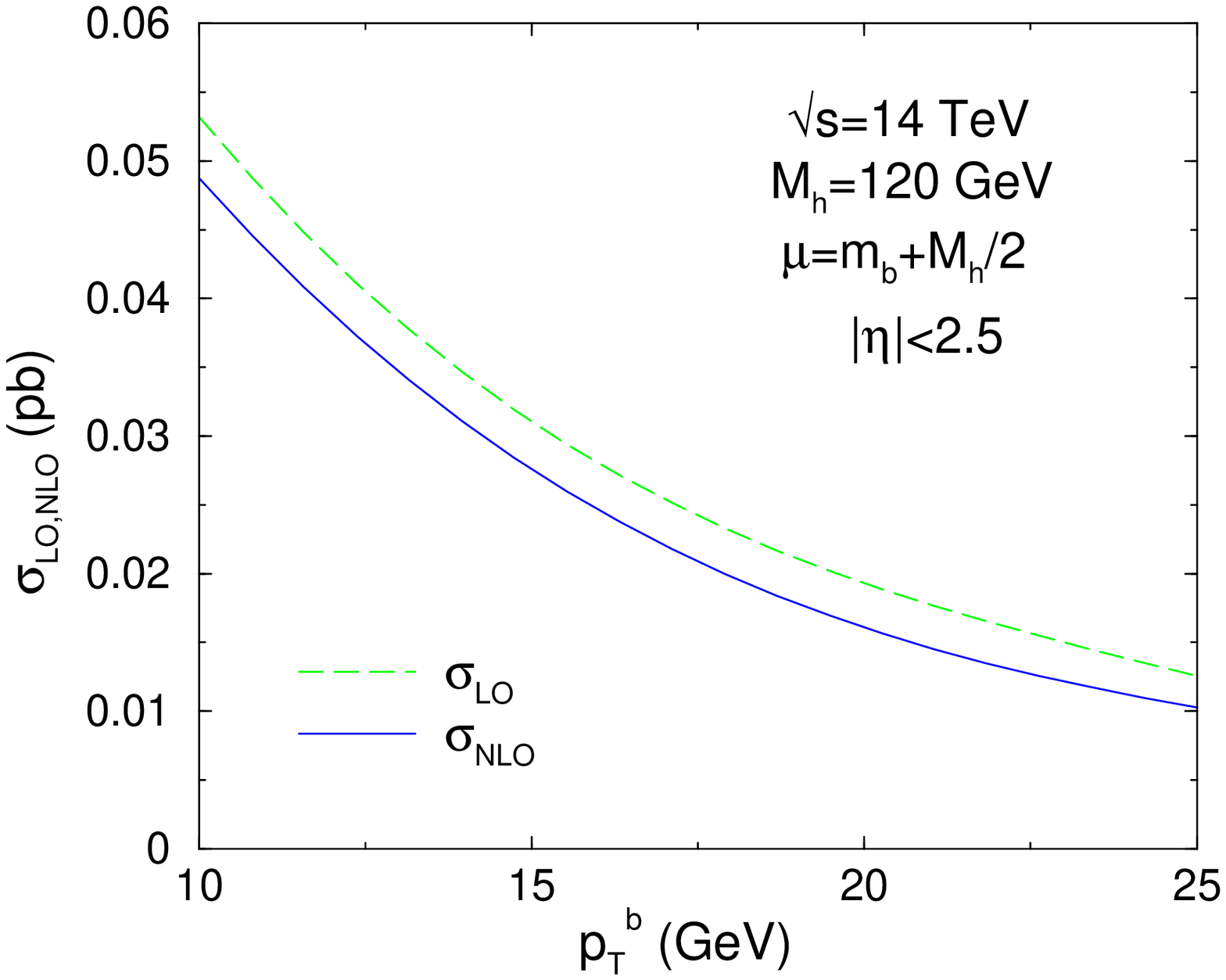} 
\caption[]{
  $\sigma_{\sss NLO,MS}$ and $\sigma_{\sss LO,MS}$ for $p\bar{p}\to
  b\bar{b}h$ at $\sqrt{s}\!=\!2$~TeV (top) and for $pp\to b\bar{b}h$
  at $\sqrt{s}\!=\!14$~TeV (bottom) as a function of the cut imposed
  on the final state bottom and anti-bottom transverse momentum
  ($p_T^b$), for $M_h\!=\!120$~GeV and $\mu\!=\!\mu_0\!=\!m_b+M_h/2$.}
\label{fg:bbh_ptcut_dep}
\end{center}
\end{figure}

Our numerical results are obtained using CTEQ5M parton distribution
functions for the calculation of the NLO cross section, and CTEQ5L
parton distribution functions for the calculation of the lowest order
cross section~\cite{Lai:1999wy}. The NLO (LO) cross section is
evaluated using the 2-loop (1-loop) evolution of $\alpha_s(\mu)$ with
$\alpha_s^{NLO}(M_Z)=0.118$. The bottom quark pole mass is taken to be
$m_b\!=\!4.6$~GeV. In the OS scheme the bottom quark Yukawa coupling
is calculated as $g_{b\bar{b}h}\!=\!m_b/v$, while in the
$\overline{MS}$ scheme as
$g_{b\bar{b}h}(\mu)\!=\!\overline{m}_b(\mu)/v$, where we use
$\overline{m}_b(\mu)_{1l}$ from Eq.~(\ref{eq:mb_msbar_rg_1l}) for
$\sigma_{\sss LO}$ and $\overline{m}_b(\mu)_{2l}$ from
Eq.~(\ref{eq:mb_msbar_rg_2l}) for $\sigma_{\sss NLO}$.

We evaluate the fully exclusive LO and NLO cross sections for
$b\bar{b}h$ production by requiring that the transverse momentum of
both final state bottom and anti-bottom quarks be larger than 20~GeV
($p_T^b\!>\!20$~GeV), and that their pseudorapidity satisfy the
condition $|\eta_b|\!<\!2$ for the Tevatron and $|\eta_b|\!<\!2.5$ for
the LHC.  This corresponds to an experiment measuring the Higgs decay
products along with two high $p_T$ bottom quark jets that are clearly
separated from the beam.  Furthermore, we present LO and NLO
transverse momentum and pseudorapidity distributions. In order to
better simulate the detector response, the gluon and the
bottom/anti-bottom quarks are treated as distinct particles only if
the separation in the azimuthal angle-pseudorapidity plane is $\Delta
R\!>\!0.4$.  For smaller values of $\Delta R$, the four momentum
vectors of the two particles are combined into an effective
bottom/anti-bottom quark momentum four-vector.

\subsection{Standard Model Results}
\label{subsec:results_sm}
 
In Fig.~\ref{fg:bbh_mu_dep} we show, for $M_h\!=\!120$~GeV, the
dependence of the LO and NLO cross sections for $p\bar{p}\to
b\bar{b}h$ at the Tevatron (top) and for $pp\to b\bar{b}h$ at the LHC
(bottom) on the unphysical factorization and renormalization scale,
$\mu$, when using either the $OS$ or the $\overline{MS}$
renormalization schemes for the bottom quark Yukawa coupling. In both
the $OS$ and $\overline{MS}$ schemes the stability of the cross
section is greatly improved at NLO, given the much milder scale
dependence with respect to the corresponding LO cross section. The
results presented in Fig.~\ref{fg:bbh_mu_dep} are obtained by setting
$\mu\!=\!\mu_r\!=\!\mu_f$, i.e. by identifying the renormalization
($\mu_r$) and factorization ($\mu_f$) scales. We have checked that
varying them independently does not affect the results significantly.
By varying the scale $\mu$ in the ranges $0.7 \mu_0<\mu<4\mu_0$
(Tevatron) and $0.5\mu_0<\mu<8\mu_0$ (LHC), when using the $OS$ scheme
for the bottom quark Yukawa coupling, and in the ranges
$0.4\mu_0<\mu<2\mu_0$ (Tevatron) and $0.2\mu_0<\mu<2\mu_0$ (LHC) when
using the $\overline{MS}$ scheme, i.~e.~in the plateau regions, the
value of the NLO cross section varies by at most 15-20\% (where
$\mu_0\!=\!m_b+M_h/2$) .

As can be seen in Fig.~\ref{fg:bbh_mu_dep}, the cross section
calculated with $g_{b\bar{b}h}$ in the $\overline{MS}$ scheme shows a
better perturbative behavior, since the difference between
$\sigma_{\sss LO}$ and $\sigma_{\sss NLO}$ is smaller. This is in part
due to the fact that the LO cross section is calculated using
$\overline{m}_b(\mu)_{1l}$ and therefore already contains some of the
corrections from the renormalization of the bottom quark Yukawa
coupling that appear in the NLO cross section as well as at higher
order. This observation seems to justify the use of
$\overline{m}_b(\mu)_{1l}$ at LO and $\overline{m}_b(\mu)_{2l}$ at
NLO. One also observes that the $\overline{MS}$ NLO cross section is
better behaved at low values of the renormalization/factorization
scales.  At the same time, both the $OS$ and $\overline{MS}$ cross
sections show well defined but distinct regions of least sensitivity
to the renormalization/factorization scale. In both cases this happens
in the region where the LO and NLO cross section are closer. The
variation of the NLO cross section with $\mu$ about its point of least
sensitivity to the renormalization/factorization scale is almost the
same whether one uses the $OS$ or $\overline{MS}$ schemes for the
bottom quark Yukawa coupling.  This indicates that the running of the
Yukawa coupling is not the only important factor to determine the
overall perturbative stability of the NLO cross section.

As discussed in Section~\ref{subsec:ren_scheme}, the numerical
difference between the two renormalization schemes can be significant.
This is illustrated in Fig.~\ref{fg:bbh_delta_mu_dep} where we plot
the absolute values of the relative difference,
$\Delta\!=\!(\sigma_{\sss NLO,OS}-\sigma_{\sss NLO,\overline{MS}})/
(\sigma_{\sss NLO,OS}+\sigma_{\sss NLO,\overline{MS}})$, between the
hadronic cross sections $\sigma_{\sss NLO,OS}$ and $\sigma_{\sss
  NLO,\overline{MS}}$ at both the Tevatron and the LHC.  As discussed
in detail at the parton level in Section~\ref{subsec:ren_scheme} (see
$\hat\Delta$ defined in Eq.(\ref{eq:diff_sigma_os_msbar})), the
difference between the two schemes is scale dependent and can be very
big for small and large scales. At the LHC, the relative difference
can be well approximated by $\Delta=\frac{\displaystyle
  1}{\displaystyle 2} A B$ with $A=\frac{\displaystyle
  \alpha_s}{\displaystyle 4\pi} g_{\sss NLO}/g_{\sss LO}$ and
$B=(1-(\overline{m}_b/m_b)^2)$, where $g_{\sss NLO,\sss LO}$
correspond to the $g_{\sss NLO,\sss LO}^{ij}$ contributions of
Eqs.~(\ref{eq:sigma_os}) and (\ref{eq:sigma_ms}) calculated at hadron
level. For instance, at $\mu=0.7 \mu_0$, $A=0.28$ and $B=0.57$, while
at $\mu=4 \mu_0$, $A=0.92$ and $B=0.66$, which shows that the
difference between the $\overline{MS}$ and the $OS$ schemes of the
bottom quark is not dominated by the running of the bottom quark mass
as it would be the case when the majority of the NLO corrections can
be absorbed in the running of $m_b$.

From both the observed similar scale dependence of $\sigma_{\sss NLO}$
in both schemes and the large numerical difference due to the
corrections that cannot be absorbed in the running of $m_b$, we
conclude that the use of the $\overline{MS}$ bottom quark Yukawa
coupling should probably not be overemphasized.  It is probably a good
approximation to take the difference between $\sigma_{\sss NLO,OS}$
and $\sigma_{\sss NLO,\overline{MS}}$ at their points of least scale
sensitivity as an upper bound on the theoretical error of the NLO
cross section, on top of the uncertainty due to the residual scale
dependence. This would amount to an additional 15-20\% uncertainty
arising from the dependence on the bottom quark Yukawa coupling
renormalization scheme.

In Fig.~\ref{fg:bbh_ptcut_dep} we illustrate the dependence of the
exclusive cross section on the $p_T$ cut imposed on the final state
bottom and anti-bottom quarks, at both the Tevatron (top) and the LHC
(bottom). We plot the LO and NLO cross sections obtained using the
$\overline{MS}$ bottom quark Yukawa coupling. Reducing the $p_T$ cut
from 25~GeV to 10~GeV approximately increases the cross section by a
factor of four. However, as the $p_T$ cut is reduced, the theoretical
calculation of the cross section becomes more unstable, because the
integration over the phase space of the final state bottom quarks
approaches more and more a region of collinear singularities.  Results
without a cut on the transverse momentum of the bottom quarks will be
presented in a later work~\cite{bbh_incl} (see also
Ref.~\cite{Dittmaier:2003ej}).

\begin{figure}[t]
\begin{center}
\includegraphics[bb=150 500 430 700,scale=0.8]{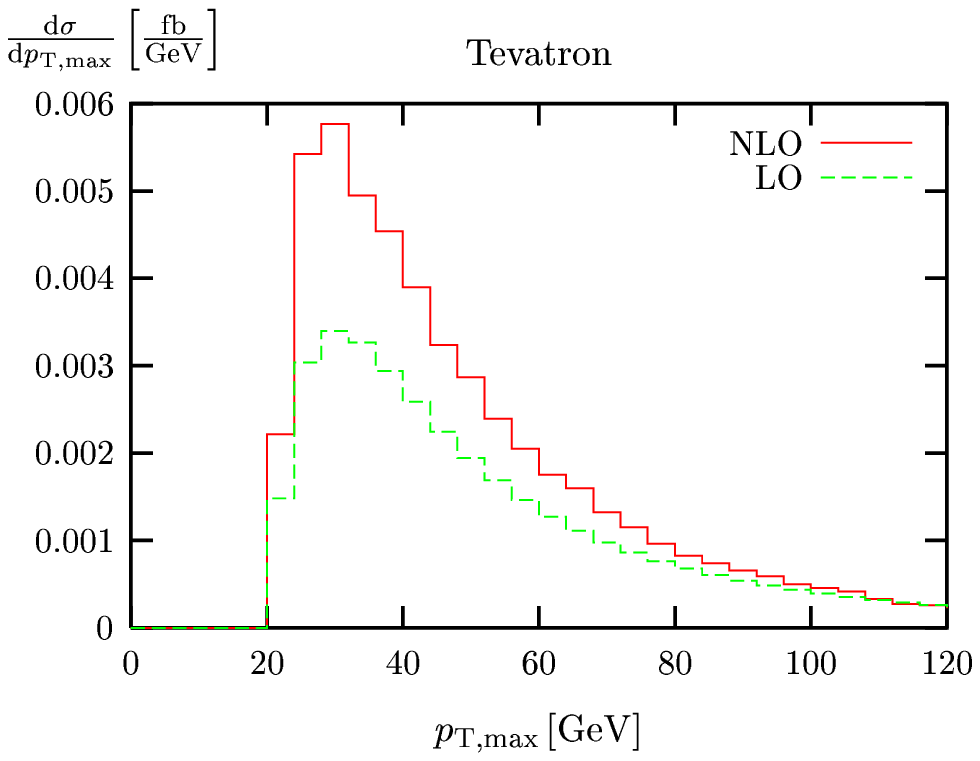}
\includegraphics[bb=150 500 430 700,scale=0.8]{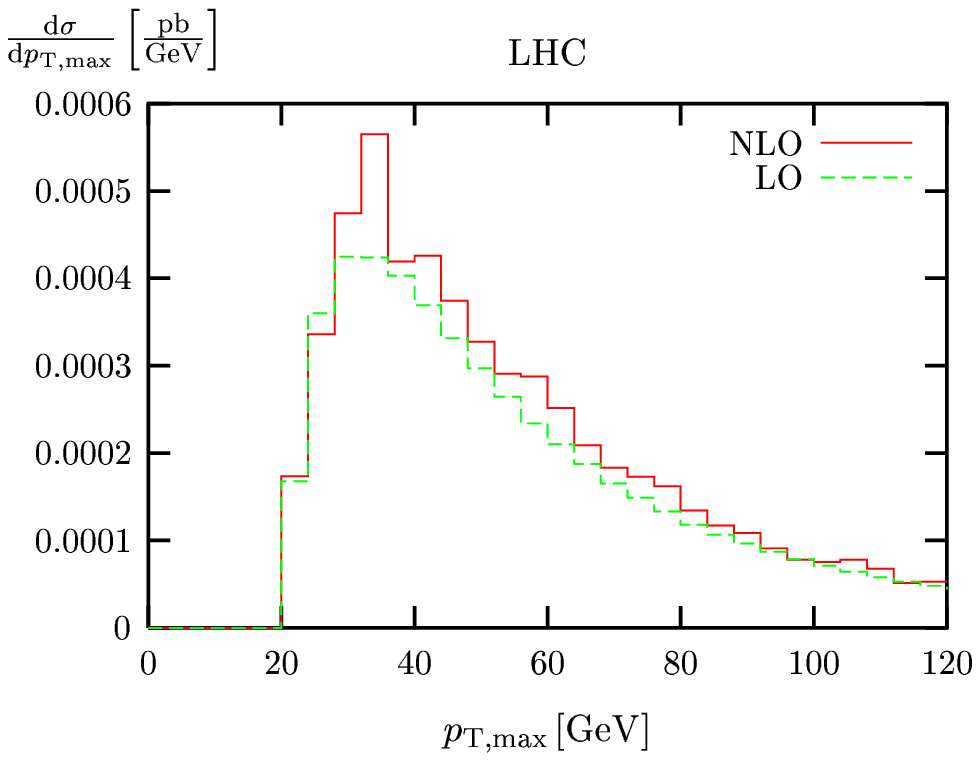} 
\caption[]{ Transverse momentum distributions at LO and NLO of the
  bottom or anti-bottom quark with the largest $p_T$. Shown are the
  $p_T^{max}$ distributions for $p\bar{p}\to b\bar{b}h$ production at
  $\sqrt{s}\!=\!2$~TeV (left) and $pp\to b\bar{b}h$ production at
  $\sqrt{s}\!=\!14$~TeV (right) in the SM and using the OS scheme for
  the bottom quark Yukawa coupling. At the Tevatron we choose $\mu\!=\!2
  m_b+M_h$, while at the LHC we choose $\mu\!=\!2 (2 m_b+M_h)$.}
\label{fg:bbh_ptmax}
\end{center}
\end{figure}

\begin{figure}[t]
\begin{center}
\includegraphics[bb=150 500 430 700,scale=0.8]{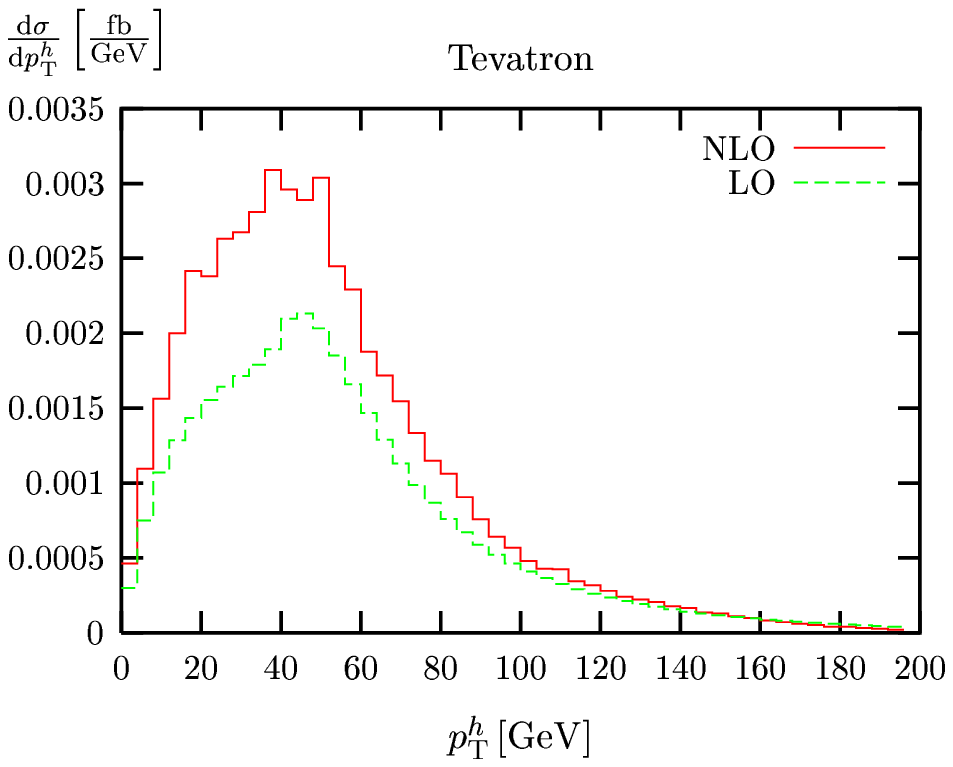} 
\includegraphics[bb=150 500 430 700,scale=0.8]{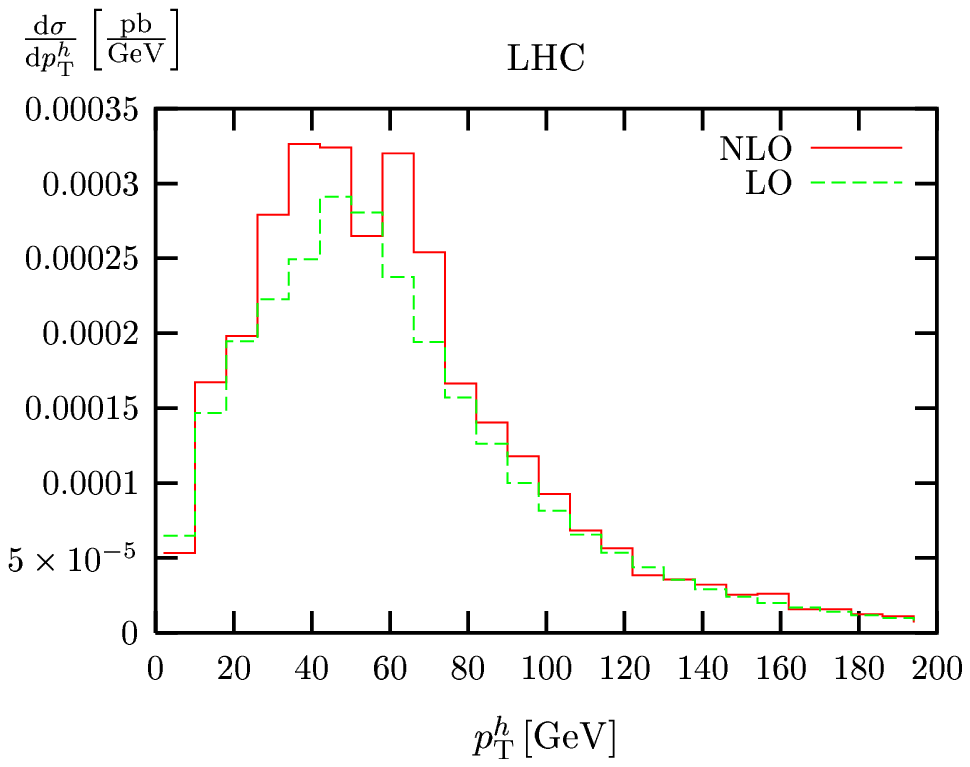} 
\caption[]{Transverse momentum distributions at LO and NLO of the
  SM Higgs boson. Shown are the $p_T^h$ distributions for $p\bar{p}\to
  b\bar{b}h$ production at $\sqrt{s}\!=\!2$~TeV (left) and $pp\to
  b\bar{b}h$ production at $\sqrt{s}\!=\!14$~TeV (right) in the SM and
  using the OS scheme for the bottom quark Yukawa coupling. At the
  Tevatron we choose $\mu\!=\!2 m_b+M_h$, while at the LHC we choose
  $\mu\!=\!2(2 m_b+M_h)$.}
\label{fg:bbh_pth}
\end{center}
\end{figure}

\begin{figure}[t]
\begin{center}
\includegraphics[bb=150 500 430 700,scale=0.8]{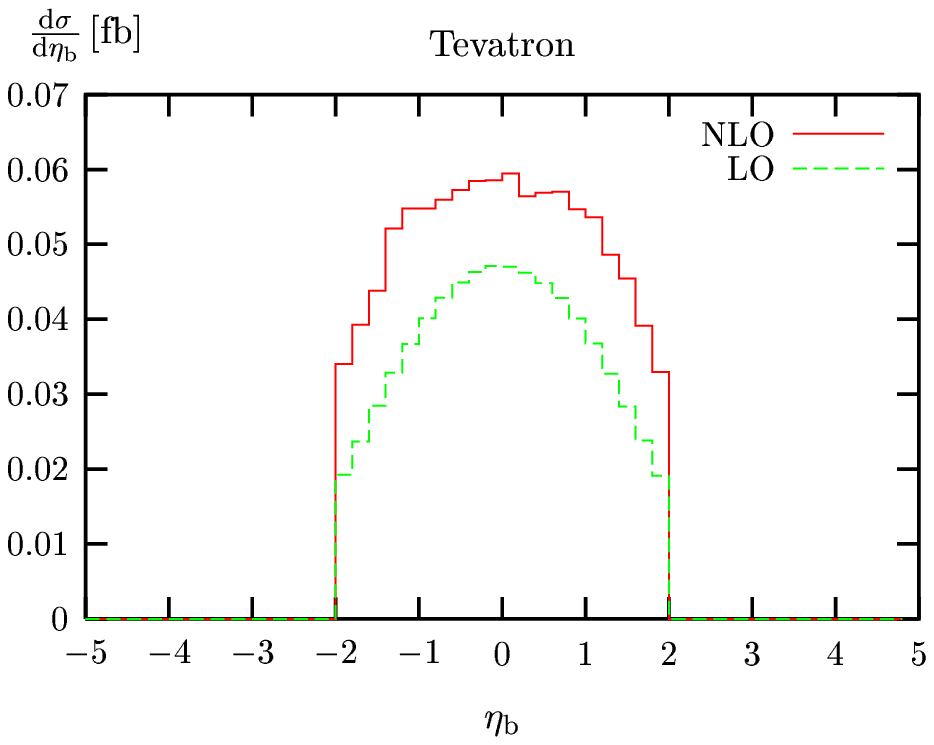} 
\includegraphics[bb=150 500 430 700,scale=0.8]{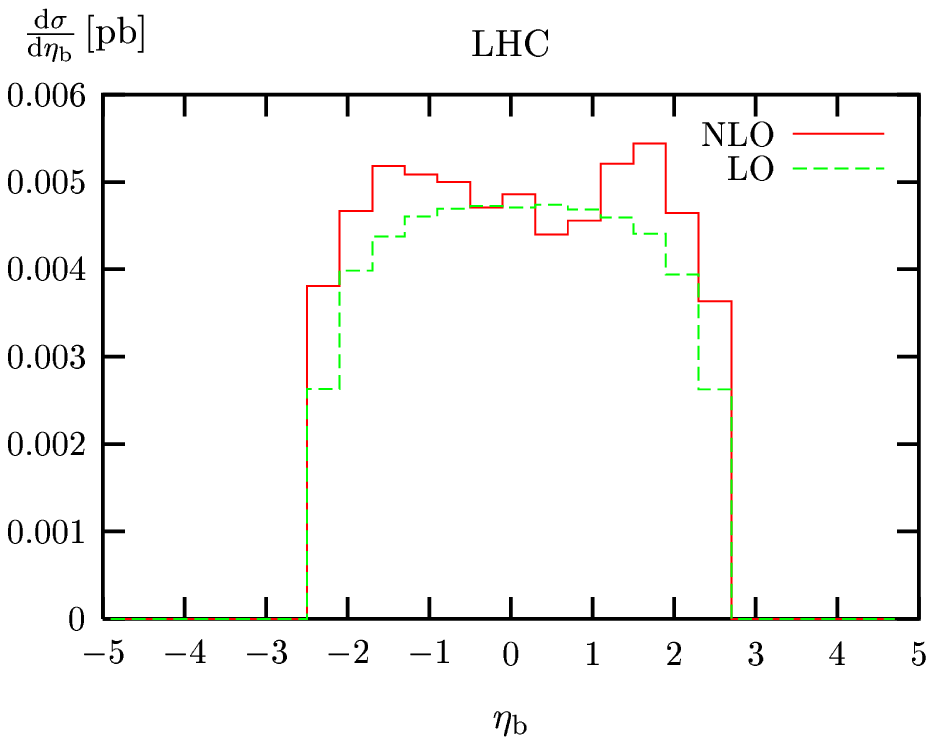} 
\caption[]{Pseudorapidity distributions at LO and NLO of the
  bottom quark. Shown are the $\eta_b$ distributions for $p\bar{p}\to
  b\bar{b}h$ production at $\sqrt{s}\!=\!2$~TeV (left) and $pp\to
  b\bar{b}h$ production at $\sqrt{s}\!=\!14$~TeV (right) in the SM and
  using the OS scheme for the bottom quark Yukawa coupling. At the
  Tevatron we choose $\mu=2 m_b+M_h$, while at the LHC we choose
  $\mu=2 (2 m_b+M_h)$.}
\label{fg:bbh_etab}
\end{center}
\end{figure}

\begin{figure}[t]
\begin{center}
\includegraphics[bb=150 500 430 700,scale=0.8]{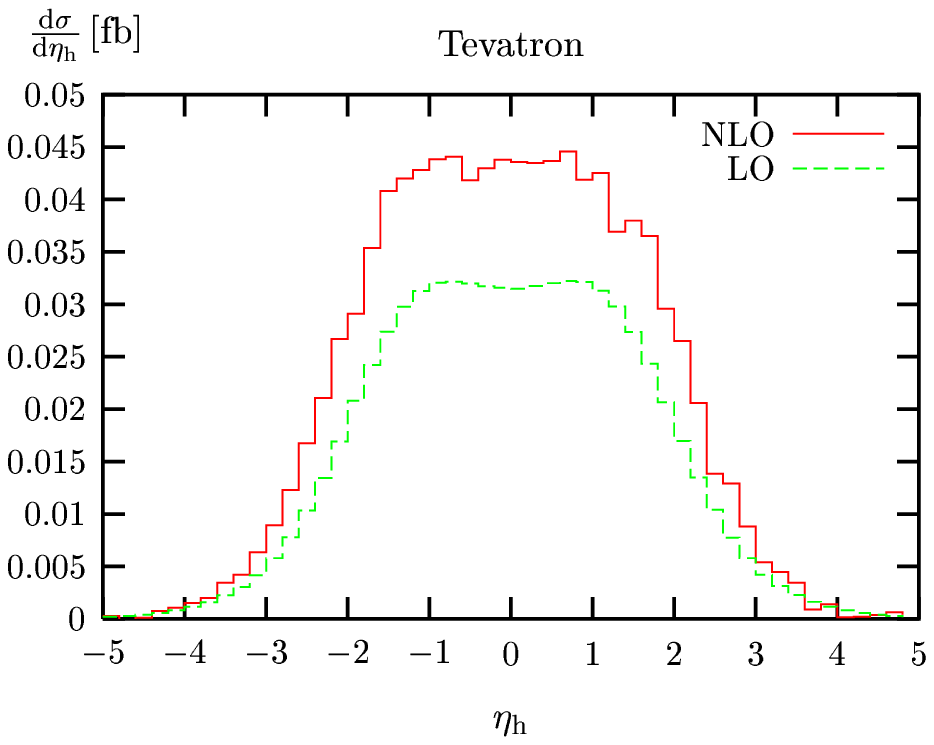} 
\includegraphics[bb=150 500 430 700,scale=0.8]{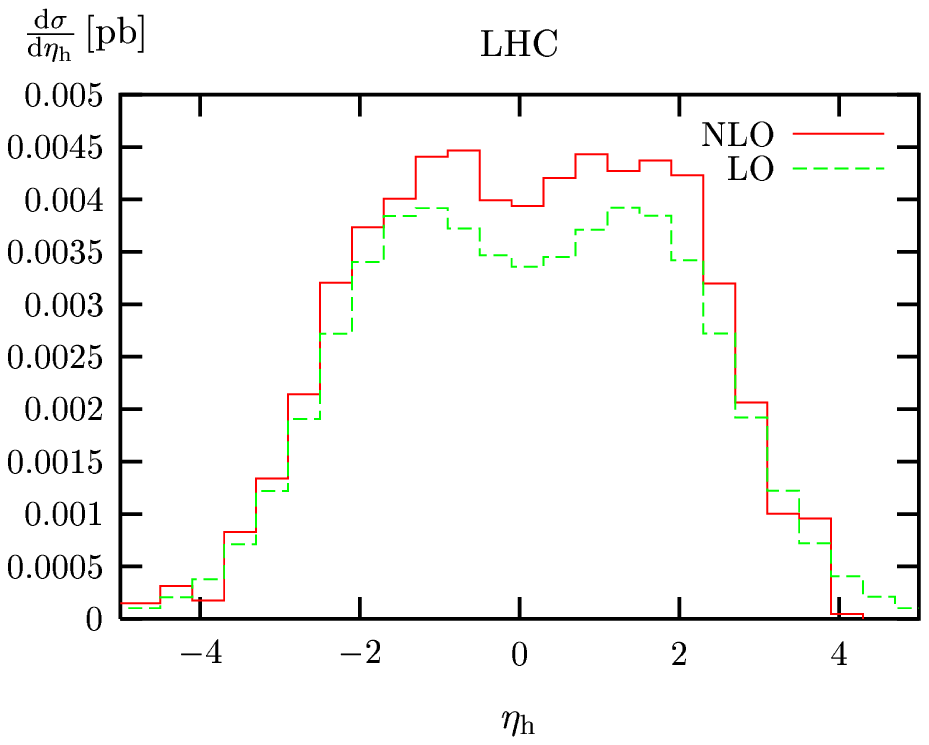} 
\caption[]{Pseudorapidity distributions at LO and NLO of the
  SM Higgs boson. Shown are the $\eta_h$ distributions for
  $p\bar{p}\to b\bar{b}h$ production at $\sqrt{s}\!=\!2$~TeV (left)
  and $pp\to b\bar{b}h$ production at $\sqrt{s}\!=\!14$~TeV (right) in
  the SM and using the OS scheme for the bottom quark Yukawa coupling.
  At the Tevatron we choose $\mu=2 m_b+M_h$, while at the LHC we
  choose $\mu=2 (2 m_b+M_h)$.}
\label{fg:bbh_etah}
\end{center}
\end{figure}

Finally, in Figs.~\ref{fg:bbh_ptmax}, \ref{fg:bbh_pth},
\ref{fg:bbh_etab}, and~\ref{fg:bbh_etah} we plot the LO and NLO
transverse momentum ($p_T$) and pseudorapidity ($\eta$) distributions
of the final state particles, the bottom and anti-bottom quarks and
the Higgs boson, both for the Tevatron and for the LHC. Both LO and
NLO differential cross sections are obtained in the SM and using the
$OS$ scheme for the bottom quark Yukawa coupling.  For the
renormalization/factorization scale we choose $\mu=2 m_b+M_h$ at the
Tevatron and $\mu=2 (2 m_b+M_h)$ at the LHC. These two scales are well
within the plateau regions where the $OS$ NLO cross sections vary the
least with the value of $\mu$. Similar results can be obtained using
the $\overline{MS}$ bottom quark Yukawa coupling.

In Fig.~\ref{fg:bbh_ptmax} we show the LO and NLO $p_T$ distributions
of the bottom or anti-bottom quark with highest $p_T$, while
Fig.~\ref{fg:bbh_pth} displays the $p_T$ distributions of the SM Higgs
boson. The pseudorapidity distributions of the bottom quark and the
Higgs boson are shown in Fig.~\ref{fg:bbh_etab} and
Fig.~\ref{fg:bbh_etah}, respectively. The inclusion of the NLO
corrections causes the cross sections to be more sharply peaked around
low $p_T^{b,h}$ and around $\eta_{b,h}\!=\!0$.

\subsection{MSSM Results}
\label{subsec:results_mssm}

\begin{figure}[t]
\begin{center}
\includegraphics[scale=0.6]{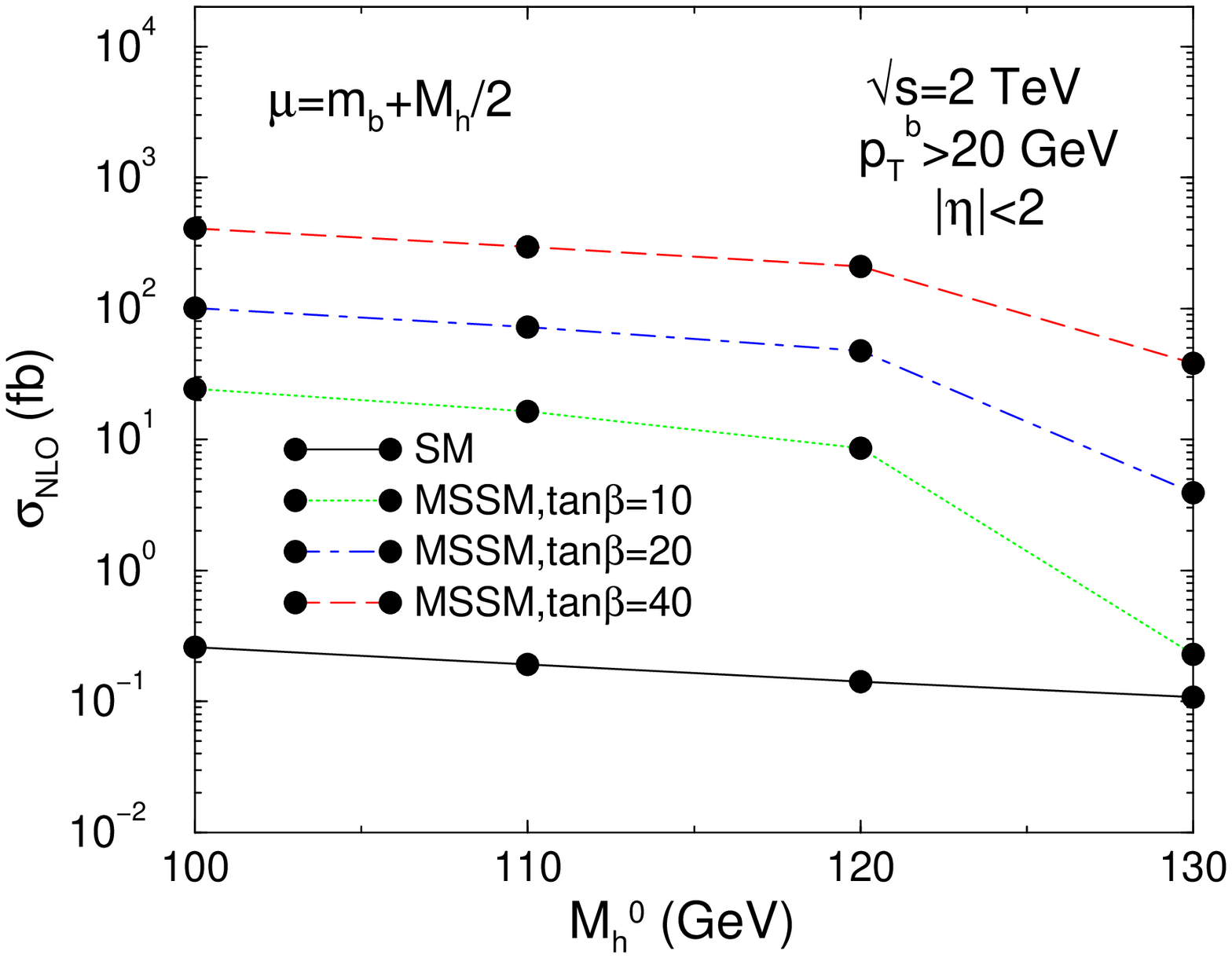} 
\includegraphics[scale=0.6]{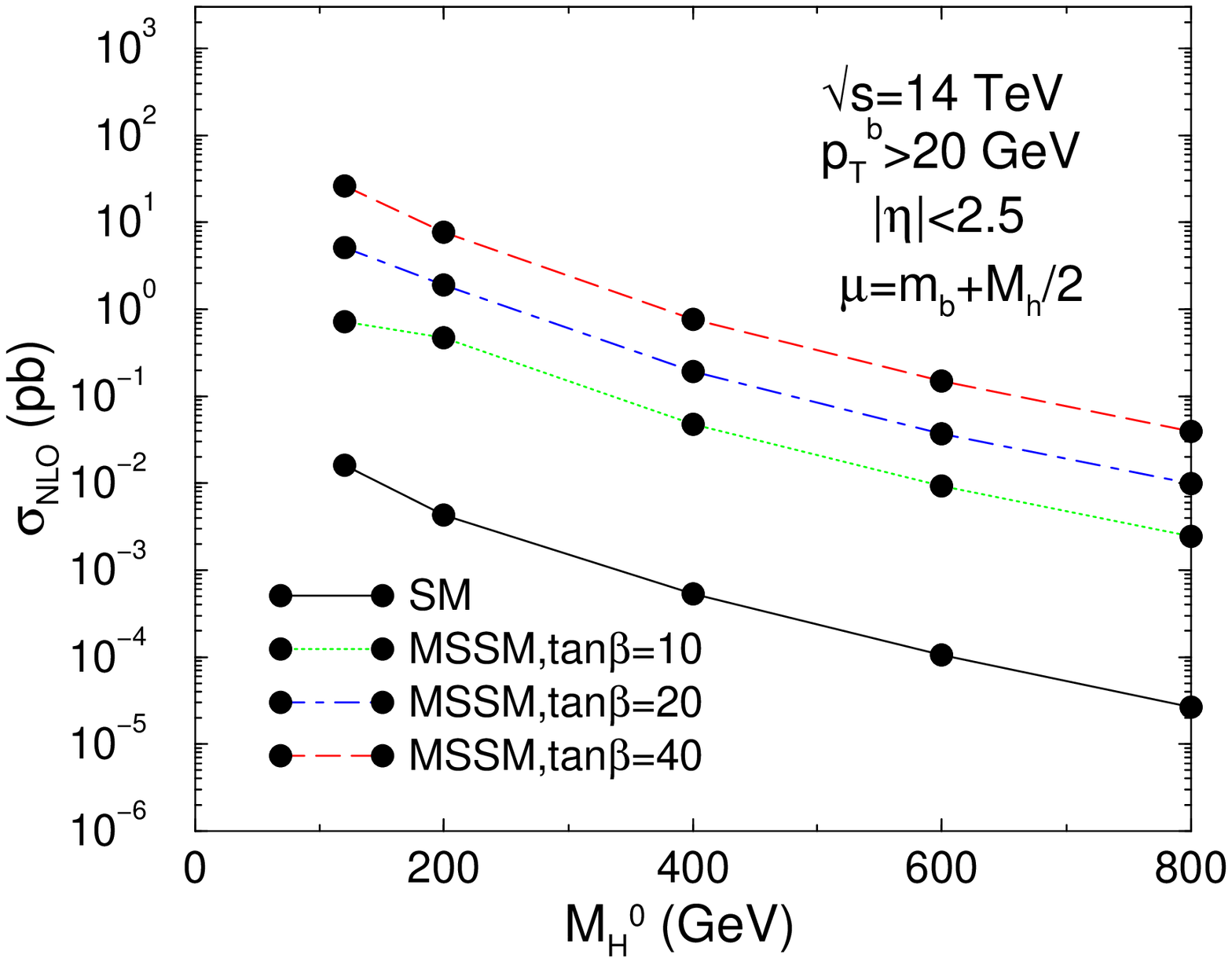} 
\caption[]{ $\sigma_{\sss NLO,MS}$ 
  for $p\bar{p}\to b\bar{b}h$ production at $\sqrt{s}\!=\!2$~TeV (top)
  and $pp\to b\bar{b}h$ production at $\sqrt{s}\!=\!14$~TeV (bottom)
  in the SM and in the MSSM with $\tan\beta\!=\!10,20$, and $40$. For
  the Tevatron we considered $p\bar{p}\to b\bar{b}h^0$ with
  $M_{h^0}\!=\!100,110,120$, and 130~GeV, while for the LHC we
  considered $pp\to b\bar{b}H^0$ with $M_{H^0}\!=\!120,200,400,600$,
  and 800~GeV. For each $(M_{h^0},\tan\beta)$ and
$(M_{H^0},\tan\beta)$ point, the corresponding values of $\alpha$
and $M_A$ are listed in Tables~\ref{tab:lighth} and \ref{tab:heavyh}.}
\label{fg:bbh_mh_dep}
\end{center}
\end{figure}

The rate for $b\bar{b}h$ production can be significantly enhanced in a
supersymmetric model with large values of $\tan\beta$. 
In the MSSM, the bottom and top quark couplings to the scalar Higgs bosons
are given by:
\renewcommand{\arraystretch}{2}
\[\begin{array}{lclrcr}
b\bar{b}h^0 & : & -\displaystyle{\frac{\sin\alpha}{\cos \beta}}g_{b\bar{b}h} 
& \; \; \; t\bar{t}h^0 & : & \; \; \displaystyle{\frac{\cos\alpha}{\sin\beta}}g_{t\bar{t}h} \\
b\bar{b}H^0 & : & \; \; \displaystyle{\frac{\cos\alpha}{\cos\beta}}g_{b\bar{b}h}
&\; \; \; t\bar{t}H^0 & : & \; \; \displaystyle{\frac{\sin\alpha}{\sin\beta}}g_{t\bar{t}h}\,\,\,
\end{array}\]
\renewcommand{\arraystretch}{1} where $g_{b\bar{b}h}$ and
$g_{t\bar{t}h}$ are the SM bottom and top quark Yukawa couplings,
$h^0$ and $H^0$ are the lighter and heavier neutral scalars of the
MSSM, and $\alpha$ is the angle which diagonalizes the neutral scalar
Higgs mass matrix~\cite{Gunion:1989we}. By replacing the SM top and
bottom quark Yukawa couplings with the corresponding MSSM ones, our
calculation can then be straightforwardly generalized to the case of
the scalar Higgs bosons of the MSSM.  The bottom quark Yukawa coupling
to the MSSM pseudoscalar Higgs boson, $A^0$, is also enhanced at large
$\tan\beta$. The corresponding cross section for $b\bar{b}A^0$
production can be obtained from our calculation in the $m_b\to 0$
limit, which we do not consider in this paper. We will present,
however, complete results for $b\bar{b}A^0$ production, i.~e.~for
non-zero $m_b$, in a future study.

The MSSM Higgs boson masses and the mixing angle $\alpha$ have been
computed up to two-loop order using the program
FeynHiggs~\cite{Heinemeyer:1998yj}.  In Tables~\ref{tab:lighth} and
\ref{tab:heavyh} we provide the values of the input parameters
($(M_{h^0},\tan\beta)$ or $(M_{H^0},\tan\beta)$) and the resulting
values of $\alpha$ used in the calculation of the top and bottom quark
Yukawa couplings to the light and heavy neutral MSSM scalar Higgs
bosons. This choice of MSSM parameters takes into account present
experimental limits on the MSSM parameter space, but represents
otherwise just one among many possible realizations of the MSSM
parameter space.  The results obtained with this choice of MSSM input
parameters illustrate the typical enhancements over the SM results one
can expect when considering the production of neutral scalar Higgs
bosons in association with bottom quarks.
    
\begin{table*}
\begin{center}
\begin{tabular}{|c|c|c|c|c|}\hline\hline
\multicolumn{5}{|c|}{$\tan\beta=10$} \\ \hline 
$M_{h^0}$ [GeV] & 100 & 110 & 120 & 130   \\ \hline 
$M_{A}$ [GeV] & 102.42 & 113.86 & 127.95 & 264.72 \\
$\alpha$ [rad]& -1.3249 & -1.1963  &  -0.9054 & -0.1463 \\ \hline
\multicolumn{5}{|c|}{$\tan\beta=20$} \\ \hline 
$M_{h^0}$ [GeV] & 100 & 110 & 120 & 130   \\ \hline 
$M_{A}$ [GeV] & 100.61 & 110.95 & 121.89  & 146.72 \\
$\alpha$ [rad]  & -1.4420 & -1.3707 &  -1.1856 & -0.3108  \\ \hline
\multicolumn{5}{|c|}{$\tan\beta=40$} \\ \hline  
$M_{h^0}$ [GeV] & 100 & 110 & 120 & 130   \\ \hline 
$M_{A}$ [GeV] & 100.15  & 110.23  & 120.46 & 133.71 \\
$\alpha$ [rad]& -1.5007 & -1.4601  & -1.3444 & -0.4999 \\ \hline \hline
\end{tabular}
\end{center}
\caption{Values of $\alpha$ and $M_A$, computed up to two-loop order by
  using the program FeynHiggs~\cite{Heinemeyer:1998yj}, corresponding to  
different choices of $\tan\beta$ and $M_{h^0}$.  In the calculation of
$\alpha$ and $M_A$ we choose the genuine SUSY input parameters as follows:
$M_{\tilde g}\!=\!M_{\tilde t_L}\!=\!M_{\tilde t_R}\!=\!
M_{\tilde b_L}\!=\!M_{\tilde b_R}\!=\!1$~TeV,
$M_{t}^{LR}\!=\!2$~TeV, $A_b\!=\!A_t\!=\!M_t^{LR}+\mu \cot\beta$, 
and $\mu\!=\!M_2\!=\!200$~GeV.}
\label{tab:lighth}
\end{table*}

The top part of Fig.~\ref{fg:bbh_mh_dep} compares the NLO $p\bar{p}\to
b\bar{b}h$ SM cross section at the Tevatron with the corresponding
cross section for production of the lightest neutral scalar Higgs
boson in the MSSM for $\tan\beta=10,20$, and $40$.  A large
enhancement of up to three orders of magnitude is observed.  As the
light neutral Higgs boson mass approaches its maximum value, the
mixing angle $\alpha$ becomes very small, as can be clearly seen in
Table~\ref{tab:lighth}. This has the effect of suppressing the $b\bar
b h^0$ rates at this point. A similar effect can be observed in the
production of a heavy neutral Higgs boson when $M_{H^0}$ is
approaching its minimum value (see Table~\ref{tab:heavyh}), as shown
in the bottom part of Fig.~\ref{fg:bbh_mh_dep}. Again, we compare the
production of the SM Higgs boson with that of the heavier neutral
scalar Higgs boson of the MSSM and observe a significant enhancement
of the rate in the MSSM for large $\tan\beta$.
\begin{table*}
\begin{center}
\begin{tabular}{|c|c|c|c|c|c|}\hline\hline
\multicolumn{6}{|c|}{$\tan\beta=10$} \\ \hline 
$M_{H^0}$ [GeV] & 120 & 200 & 400 & 600 & 800   \\ \hline 
$M_{A}$ [GeV] & 108.05  & 198.55 & 399.41 & 599.64   & 799.74 \\
$\alpha$ [rad]& -0.9018 & -0.1762  & -0.1140 & -0.1057 & -0.1030 \\ \hline
\multicolumn{6}{|c|}{$\tan\beta=20$} \\ \hline  
$M_{H^0}$ [GeV] & 120 & 200 & 400 & 600 & 800   \\ \hline 
$M_{A}$ [GeV] & 116.45 & 199.56 & 399.81  & 599.89 & 799.91 \\
$\alpha$ [rad]& -0.5785 & -0.0901 & -0.0574  & -0.0531 & -0.0517 \\ \hline
\multicolumn{6}{|c|}{$\tan\beta=40$} \\ \hline  
$M_{H^0}$ [GeV] & 120 & 200 & 400 & 600 & 800   \\ \hline 
$M_{A}$ [GeV] & 118.92 & 199.82 & 399.92 & 599.95 & 799.96 \\
$\alpha$ [rad]& -0.3116 & -0.0460 & -0.0289  & -0.0267 & -0.0259 \\ \hline \hline
\end{tabular}
\end{center}
\caption{\label{tab:heavyh}
Values of $\alpha$ and $M_A$, computed up to two-loop order by using the program
FeynHiggs~\cite{Heinemeyer:1998yj}, corresponding to different choices of 
$\tan\beta$ and $M_{H^0}$. In the calculation of $\alpha$ and $M_A$ 
we choose the genuine SUSY input parameters as follows:
$M_{\tilde g}=M_{\tilde t_L}=M_{\tilde t_R}
=M_{\tilde b_L}=M_{\tilde b_R}=1$~TeV,
$M_{t}^{LR}=0$, $A_b=A_t=M_t^{LR}+\mu \cot\beta$, and $\mu=M_2=1$~TeV.
}
\end{table*}

\section{Conclusions}
\label{sec:conclusions}

We presented results for the next-to-leading order QCD cross section
for exclusive $b\bar{b}h$ production at both the Tevatron and the LHC.
Our NLO results show an improved stability with respect to the
unphysical factorization and renormalization scales as compared to the
leading order results and increase the reliability of the theoretical
prediction. The uncertainty in the resummation of large logarithms
from higher order corrections, however, is also visible in the
dependence of the NLO cross section on the renormalization scheme of
the bottom quark Yukawa coupling. The residual
renormalization/factorization scale dependence is of the order of
15-20\% when the bottom quark Yukawa coupling is renormalized in the
$OS$ or $\overline{MS}$ schemes respectively. We conservatively
estimate the additional uncertainty due to the renormalization scheme
dependence of the bottom quark Yukawa coupling to be at most of order
15-20\%.

Our calculation is important for Higgs boson searches at hadron
colliders where two high $p_T$ bottom quarks are tagged in the final
state.  In supersymmetric models with large $\tan\beta$, $b\bar{b}h$
production can be an important discovery channel, at both the Tevatron
and the LHC.

\section*{Acknowledgments}
We thank R.~Harlander, M.~Kr\"amer, J.~K\"uhn, F.~Maltoni, and S.
Willenbrock for valuable discussions. S.D. and L.R. would like to
thank the organizers of the Les Houches Workshop on \emph{Physics at
  TeV Colliders} for providing such a pleasant and stimulating
environment where many of the issues presented in this paper were
extensively discussed. L.R. acknowledges the kind hospitality of the
Theory Division at CERN and of the Particle Physics group of the IST
in Lisbon while part of this work was being completed. The work of
S.D. (C.B.J. and L.R.)  is supported in part by the U.S. Department of
Energy under grant DE-AC02-98CH10886 (DE-FG02-97ER41022). The work of
D.W. is supported in part by the National Science Foundation under
grant No.~PHY-0244875.

\bibliography{bbh}
\end{document}